\documentclass[preprint2,letteredappendix,appendixfloats]{aastex631}

\usepackage{amsmath}
\usepackage{romannum}
\usepackage{natbib}
\usepackage{enumitem}
\usepackage{soul}
\usepackage{longtable}

\usepackage{graphicx}           % for including figures
\usepackage{latexsym}           % additional math symbols
\usepackage{bm}                 % bold math package
\usepackage[english]{babel}
\usepackage{color}              % for colored text and boxes; remove for subm.

\begin{document}
\pagenumbering{arabic}

\title{JWST view of four infant galaxies at \boldsymbol{$z=8.31$}--8.49 in the MACS0416 field 
and implications for reionization}
\shorttitle{Infant galaxies in the MACS0416 field}
\shortauthors{Ma, Sun, Cheng, et al.}

\correspondingauthor{Haojing Yan}
\email{yanha@missouri.edu}

\author[0000-0003-3270-6844]{Zhiyuan Ma}
\affiliation{Department of Astronomy, University of Massachusetts, Amherst, USA, 01003}

\author[0000-0001-7957-6202]{Bangzheng Sun}
\affiliation{Department of Physics and Astronomy, University of Missouri, Columbia, MO\,65211, USA}

\author[0000-0003-0202-0534]{Cheng Cheng}
\affiliation{Chinese Academy of Sciences South America Center for Astronomy, National Astronomical Observatories, CAS, Beijing 100101, China}

\author[0000-0001-7592-7714]{Haojing Yan}
\affiliation{Department of Physics and Astronomy, University of Missouri, Columbia, MO\,65211, USA}

\author[0000-0003-4952-3008]{Chenxiaoji Ling}
\affiliation{National Astronomical Observatories, Chinese Academy of Science, Beijing, 100101, China}

\author[0000-0002-4622-6617]{Fengwu Sun}
\affiliation{Center for Astrophysics \textbar\ Harvard \& Smithsonian, 60 Garden 
Street, Cambridge, MA 02138, USA}
\affiliation{Steward Observatory, University of Arizona, 933 N Cherry Ave, Tucson, AZ, 85721-0009, USA}

\author[0000-0002-7460-8460]{Nicholas Foo}
\affiliation{School of Earth and Space Exploration, Arizona State University,
Tempe, AZ 85287-1404, USA}

\author[0000-0003-1344-9475]{Eiichi Egami}
\affiliation{Steward Observatory, University of Arizona, 933 N Cherry Ave, Tucson, AZ, 85721-0009, USA}

\author[0000-0001-9065-3926]{Jos\'e M. Diego}
\affiliation{Instituto de Física de Cantabria (CSIC-UC), Avda. Los Castros s/n, 39005, Santander, Spain}

\author[0000-0003-3329-1337]{Seth H. Cohen} %%% seth.cohen@asu.edu
\affiliation{School of Earth and Space Exploration, Arizona State University,
Tempe, AZ 85287-1404, USA}

\author[0000-0003-1268-5230]{Rolf A. Jansen} %%% rolfjansen.work@gmail.com
\affiliation{School of Earth and Space Exploration, Arizona State University,
Tempe, AZ 85287-1404, USA}

\author[0000-0002-7265-7920]{Jake Summers} %%% jssumme1@asu.edu
\affiliation{School of Earth and Space Exploration, Arizona State University,
Tempe, AZ 85287-1404, USA}

\author[0000-0001-8156-6281]{Rogier A. Windhorst}%%% Rogier.Windhorst@gmail.com
\affiliation{School of Earth and Space Exploration, Arizona State University,
Tempe, AZ 85287-1404, USA}

\author[0000-0002-9816-1931]{Jordan C. J. D'Silva} %%% jordan.dsilva@research.uwa.edu.au
\affiliation{International Centre for Radio Astronomy Research (ICRAR) and the
International Space Centre (ISC), The University of Western Australia, M468,
35 Stirling Highway, Crawley, WA 6009, Australia}
\affiliation{ARC Centre of Excellence for All Sky Astrophysics in 3 Dimensions
(ASTRO 3D), Australia}

\author[0000-0002-6610-2048]{Anton M. Koekemoer} %%% koekemoer@stsci.edu
\affiliation{Space Telescope Science Institute,
3700 San Martin Drive, Baltimore, MD 21218, USA}

\author[0000-0001-7410-7669]{Dan Coe} %%% dcoe@stsci.edu
\affiliation{Space Telescope Science Institute, 3700 San Martin Drive, Baltimore, MD 21218, USA}
\affiliation{Association of Universities for Research in Astronomy (AURA) for the European Space Agency (ESA), STScI, Baltimore, MD 21218, USA}
\affiliation{Center for Astrophysical Sciences, Department of Physics and Astronomy, The Johns Hopkins University, 3400 N Charles St. Baltimore, MD 21218, USA}

\author[0000-0003-1949-7638]{Christopher J. Conselice} %%% conselice@gmail.com
\affiliation{Jodrell Bank Centre for Astrophysics, Alan Turing Building,
University of Manchester, Oxford Road, Manchester M13 9PL, UK}

\author[0000-0001-9491-7327]{Simon P. Driver} %%% Simon.Driver@icrar.org
\affiliation{International Centre for Radio Astronomy Research (ICRAR) and the
International Space Centre (ISC), The University of Western Australia, M468,
35 Stirling Highway, Crawley, WA 6009, Australia}

\author[0000-0003-1625-8009]{Brenda Frye} %%% brendafrye@gmail.com
\affiliation{Department of Astronomy/Steward Observatory, University of Arizona, 933 N Cherry Ave,
Tucson, AZ, 85721-0009, USA}

\author[0000-0001-9440-8872]{Norman A. Grogin} %%% nagrogin@stsci.edu
\affiliation{Space Telescope Science Institute,
3700 San Martin Drive, Baltimore, MD 21218, USA}

\author[0000-0001-6434-7845]{Madeline A. Marshall} %%% madeline_marshall@outlook.com
\affiliation{National Research Council of Canada, Herzberg Astronomy \&
Astrophysics Research Centre, 5071 West Saanich Road, Victoria, BC V9E 2E7,
Canada}
\affiliation{ARC Centre of Excellence for All Sky Astrophysics in 3 Dimensions
(ASTRO 3D), Australia}

\author[0000-0001-6342-9662]{Mario Nonino} %%% nnn.mario@gmail.com
\affiliation{INAF-Osservatorio Astronomico di Trieste, Via Bazzoni 2, 34124
Trieste, Italy} %%% Mario will continue to be builders co-author, see ackn below

\author[0000-0002-6150-833X]{Rafael {Ortiz~III}} %%% rortizii@asu.edu
\affiliation{School of Earth and Space Exploration, Arizona State University,
Tempe, AZ 85287-1404, USA}

\author[0000-0003-3382-5941]{Nor Pirzkal} %%% npirzkal@stsci.edu
\affiliation{Space Telescope Science Institute,
3700 San Martin Drive, Baltimore, MD 21218, USA}

\author[0000-0003-0429-3579]{Aaron Robotham} %%% aaron.robotham@uwa.edu.au
\affiliation{International Centre for Radio Astronomy Research (ICRAR) and the
International Space Centre (ISC), The University of Western Australia, M468,
35 Stirling Highway, Crawley, WA 6009, Australia}

\author[0000-0003-0894-1588]{Russell E. Ryan, Jr.} %%% rryan.asu@stsci.edu
\affiliation{Space Telescope Science Institute,
3700 San Martin Drive, Baltimore, MD 21218, USA}

\author[0000-0001-9262-9997]{Christopher N. A. Willmer} %%% cnawillmer@gmail.com
\affiliation{Steward Observatory, University of Arizona,
933 N Cherry Ave, Tucson, AZ, 85721-0009, USA}

\author[0000-0003-4875-6272]{Nathan J. Adams} %%% axemannatbot@gmail.com 
\affiliation{Jodrell Bank Centre for Astrophysics, Alan Turing Building, 
University of Manchester, Oxford Road, Manchester M13 9PL, UK}

\author[0000-0001-6145-5090]{Nimish P. Hathi}
\affiliation{Space Telescope Science Institute, 3700 San Martin Drive,
Baltimore, MD 21218, USA}

\author[0000-0002-9767-3839]{Herv\'{e} Dole}
\affiliation{Universit\'e Paris-Saclay, CNRS, Institut d'Astrophysique Spatiale, 
91405, Orsay, France}

\author[0000-0002-9895-5758]{S. P. Willner}
\affiliation{Center for Astrophysics \textbar\ Harvard \& Smithsonian, 60 Garden 
Street, Cambridge, MA 02138, USA}

\author[0000-0002-8726-7685]{Daniel Espada}
\affiliation{Departamento de F\'{i}sica Te\'{o}rica y del Cosmos, Campus de Fuentenueva, Edificio Mecenas, Universidad de Granada, E-18071, Granada, Spain}
\affiliation{Instituto Carlos I de F\'{i}sica Te\'{o}rica y Computacional, Facultad de Ciencias, E-18071, Granada, Spain}

\author[0000-0001-6278-032X]{Lukas J. Furtak}
\affiliation{Physics Department, Ben-Gurion University of the Negev, P.O. Box 653, Beer-Sheva 8410501, Israel}

\author[0000-0003-4512-8705]{Tiger Yu-Yang Hsiao}
\affiliation{Center for Astrophysical Sciences, Department of Physics and Astronomy, The Johns Hopkins University, 3400 N Charles St., Baltimore, MD 21218, USA}
\affiliation{Space Telescope Science Institute,
3700 San Martin Drive, Baltimore, MD 21218, USA}

\author[0000-0002-3119-9003]{Qiong Li} %qiong.li@manchester.ac.uk
\affiliation{Jodrell Bank Centre for Astrophysics, Alan Turing Building,
University of Manchester, Oxford Road, Manchester M13 9PL, UK}

\author[0000-0003-1060-0723]{Wenlei Chen}
\affiliation{Department of Physics, Oklahoma State University, 145 Physical Sciences Bldg, Stillwater, OK 74078, USA}

\author[0000-0002-3405-5646]{Jean-Baptiste Jolly}
\affiliation{Max-Planck-Institut für Extraterrestrische Physik (MPE), Giessenbachstraße 1, D-85748 Garching, Germany}

\author[0000-0002-3805-0789]{Chian-Chou Chen}
\affiliation{Academia Sinica Institute of Astronomy and Astrophysics (ASIAA), No. 1, Section 4, Roosevelt Road, Taipei 10617, Taiwan}

\begin{abstract}

{New JWST/NIRCam wide-field slitless spectroscopy provides redshifts for four 
$z>8$ galaxies located behind the lensing cluster MACS J0416.1$-$2403.   
Two of them, ``Y1'' and ``JD'', have previously reported spectroscopic redshifts
based on ALMA measurements of [\ion{O}{3}]~88~$\mu$m and/or
[\ion{C}{2}]~157.7~$\mu$m lines. Y1 is a merging system of three
components, and the existing redshift $z=8.31$ is confirmed. However, JD is at 
$z=8.34$ instead of the previously claimed $z=9.28$. JD's close companion, 
``JD-N'', which was a previously discovered $z>8$ candidate, is now identified 
at the same redshift as JD\null. JD and JD-N form an
interacting pair. A new candidate at $z>8$, ``f090d\_018'', is also confirmed 
and is at $z=8.49$. These four objects are likely part of an overdensity that 
signposts a large structure extending $\sim$165~kpc in projected distance and
$\sim$48.7~Mpc in radial distance. They are magnified by less than one magnitude and have 
intrinsic $M_{\rm UV}$ ranging from $-19.57$ to $-20.83$~mag. Their spectral 
energy distributions show that the galaxies are all very young with 
ages $\sim$4--18~Myr and stellar masses about $10^{7-8}$~${\rm M_\odot}$. These 
infant galaxies have very different star formation rates ranging from a few to 
over a hundred $\rm{M_\odot}$~yr$^{-1}$, but only two of them (JD and
f090d\_018) have blue rest-frame UV slopes $\beta<-2.0$ indicative of 
a high Lyman-continuum photon escape fraction that could contribute 
significantly to the cosmic hydrogen-reionizing background. Interestingly, these
two galaxies are the least massive and least active ones among the four. The 
other two systems have much flatter UV slopes largely because of their high dust 
extinction ($A_{\rm V}=0.9$--1.0~mag).
Their much lower indicated escape fractions show that even very young, actively 
star-forming galaxies can have negligible contribution to reionization when they
quickly form dust throughout their bodies.
}

\end{abstract}
%\keywords{High Redshift --- Near Infrared}

\section{Introduction} \label{sec:intro}

    Young, star-forming galaxies are long thought to be the major drivers of the
cosmic hydrogen reionization because they should have strong UV emission and are
sufficiently abundant at $z>6$ 
\citep[e.g.,][]{Yan2004a, Bouwens2006, Finkelstein2012, Robertson2022, Atek2024}.
Their exact contribution to the ionizing photon background, however, still 
depends on how effectively their Lyman continuum (LyC; $\lambda<912$\AA) 
photons can escape and reach the surrounding intergalactic medium (IGM). Direct 
measurement of this escape fraction ($f_{\rm esc}$) for star-forming galaxies 
at $z>6$ is impossible, because the line-of-sight IGM H~\Romannum{1} absorption 
at such redshifts wipes out any LyC photons. As the alternative, $f_{\rm esc}$ 
has been measured for the ``analogs'' at lower redshifts ($z\lesssim 3$--4) 
where the IGM hydrogen is fully ionized 
\citep[e.g.][]{Steidel2001, Steidel2018, Vanzella2012, Siana2015, Grazian2016, Izotov2016, Izotov2018, Izotov2021, SL2022, griffiths2022, Flury2022, Citro2024}, 
and the correlations between $f_{\rm esc}$ and various observables are sought 
after in order to provide some viable routes to indirectly measure 
$f_{\rm esc}$ at $z>6$. Among all possible correlations, the one with the 
rest-frame UV slope (commonly denoted as $\beta$; $f_\lambda\propto \lambda^{\beta}$) is the most promising 
\citep[e.g.,][]{Zackrisson2013, Chisholm2022}.
    
    There have been ample studies of the UV slopes of galaxies at $z>6$ using
the Hubble Space Telescope (HST) deep survey data
\citep[e.g.,][]{Hathi2008, Bouwens2010, Dunlop2012, Dunlop2013, Finkelstein2012},
and the investigation has been extended to higher redshifts and fainter limits
by the James Webb Space Telescope 
\citep[JWST; e.g.,][]{Topping2022, Cullen2023, Nanayakkara2023, Morales2024, Austin2024}.
However, the vast majority of those are based on photometric samples of 
candidates that have not yet been spectroscopically confirmed. Thanks to the 
growing number of JWST spectroscopic programs, the situation is now  quickly 
changing
\citep[e.g.,][]{Tang2023, Fujimoto2023a, Saxena2024, RB2024}. 
Nevertheless, there are 
still few studies using confirmed $z>6$ galaxies. In particular, there seems to 
be a lack of extremely young galaxies (age $\lesssim 30$~Myr) in the samples. 
Such galaxies presumably should have the bluest UV slopes because their UV emission must be dominated by O and B stars.

    This work presents a case study using JWST data of {four $z\approx8.3$--8.5} galaxies behind the MACS J0416.1$-$2403 cluster (hereafter ``MACS0416''). 
MACS0416 is one of the six Hubble Frontier Fields \citep[HFF;][]{Lotz2017} and 
one of the targets of the Reionization Lensing Cluster Survey 
\citep[RELICS;][]{Coe2019}. A few investigations have used HST to search for 
lensed high-redshift galaxies behind this cluster 
\citep[e.g.,][]{Coe2015, Infante2015, Laporte2015}, and some candidates with 
photometric redshifts $z_{\rm ph}\gtrsim 8$--9 have been found. Among them, two 
have reported spectroscopic redshifts at $z>8$, both based on ALMA spectroscopy. 
One is MACSJ0416.1\_Y1 (hereafter ``Y1'' for simplicity), which was first 
selected as a $z\approx 8$ candidate
{\citep[][$z_{\rm ph}=8.1$--8.9 with best-fit value 8.57]{Laporte2015}}
and was later confirmed at $z=8.31$ 
through the detections of the [\ion{O}{3}]~88~$\mu$m 
line \citep{Tamura2019} and the [\ion{C}{2}]~157.7~$\mu$m line
\citep{Bakx2020}. The other is MACS0416.1-JD (hereafter ``JD''), which was also 
first discovered as a $z\approx 8$ candidate by 
\citet[][object ``FFC2-1151-4540'']{Coe2015} 
{with $z_{\rm ph}=7.3$--8.6 and 
best-fit $z_{\rm ph}= 8.1$. }
JD was independently rediscovered by \citet[][their object ``MACSJ0416.1\_Y2'']
{Laporte2015} 
{with $z_{\rm ph}=8.3$--8.6 and  best-fit $z_{\rm ph}= 8.47$.
(See also \citealt{Infante2015, Laporte2016}.)
\citet[][]{Laporte2021MNRAS} reported detection of the 
[\ion{O}{3}]~88~$\mu$m line at $z=9.28$ and renamed this object to 
``JD'', which we have adopted for brevity.
}
As we will show, the new 
JWST data confirm the redshift of Y1 at $z=8.31$ but change that for JD to 
$z=8.34$.
In addition, the close neighbor to JD, object ``MACSJ0416.1\_Y3'' 
as reported by
{\citet[][$z_{\rm ph}=8.8$--9.7 with best-fit $z_{\rm ph}= 9.29$]{Laporte2015}, 
}
is also at $z=8.34$. 
{Furthermore, we have identified a new JWST NIRCam 
F090W dropout, which we name f090d\_018. It turns out to have a similar redshift, $z=8.49$.}
Interestingly, all four objects have very young ages ($\sim$4--18~Myr) inferred from 
their spectral energy distributions (SEDs), and yet only two of them
show blue UV slopes of $\beta< -2.0$.

   This paper is organized as follows. Section 2 describes the JWST 
data, and Section 3 presents the photometric and spectroscopic results. The 
analysis is given in Section 4, and Section~5 is a brief summary. We adopt a 
flat $\Lambda$ cold-dark-matter cosmology with $h=0.7$,  $\Omega_{\Lambda}=0.7$, and $\Omega_M=0.3$. All magnitudes are in the AB system, 
and all coordinates are in the ICRS frame (equinox 2000).

\section{JWST NIRCam Observations and Data Reduction} \label{sec:data}

\begin{figure*}
    \centering
    \includegraphics[width=\textwidth,height=\textheight,keepaspectratio]{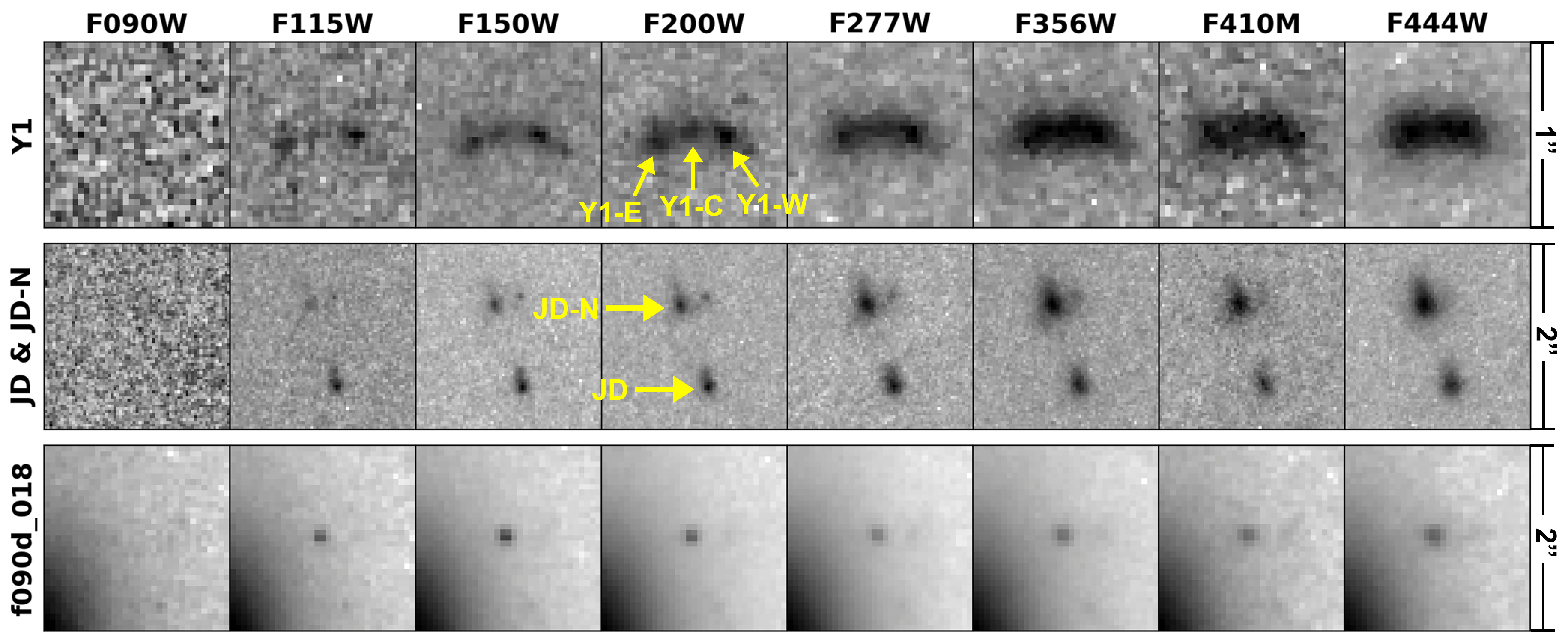}
    \caption{Cutouts of Y1 (top), JD (middle), and f090d\_018 (bottom) in the eight NIRCam
    bands using the 30\,mas images. North is up and East is left.
    The size of the Y1 images is 
    1\arcsec$\times$1\arcsec, while others are
    2\arcsec$\times$2\arcsec. The three resolved Y1 components
    (Y1-E, Y1-C, and Y1-W) are labeled in the F200W image. JD has a close 
    companion, JD-N, whose colors also meet the requirements of a $J$-band dropout. Both objects are labeled in the F200W image. 
    }
    \label{fig:nircamimg}
\end{figure*}

\subsection{NIRCam Imaging}
 
    MACS0416 is one of the targets observed by the JWST GTO program Prime 
Extragalactic Areas for Reionization and Lensing Science (PEARLS; 
PI: R. Windhorst; PID 1176; \citealt{Windhorst2023}). It was imaged by NIRCam 
in eight filters, F090W, F115W, F150W, and F200W in the short 
wavelength channel (SW) and F277W, F356W, F410M, and F444W in the long 
wavelength channel (LW)\null. The observations were  carried out in three epochs: 2022 Oct 7 (Ep1), 2022 Dec 29 (Ep2), and 2023 Feb 
10 (Ep3). The observations were described by 
\citet[][their Table~1]{Yan2023c}, who gave details of the data reduction. 
In \citet{Yan2023c}, the images were stacked on a per-epoch basis for the 
transient studies. For this work, we combined the 
data in all three epochs, which reached the total integration times of 
8761 seconds in F150W, F200W, F277W, and F356W, 10909 seconds in F115W and 
F410M, and 11338 seconds in F090W and F444W\null. In Ep3, Y1 is 
contaminated by a strong diffraction spike of a bright star, and therefore we 
only combined Ep1 and Ep2 data for the study of this object. The total 
integration times are 5841 seconds in F150W, F200W, F277W, and F356W, and 7559 
seconds in F090W, F115W, F410M, and F444W, respectively. As  
\citet{Yan2023c}, we created stacks at scales of both 0\farcs06 
(``60\,mas'') and 0\farcs03 (``30\,mas''), which have magnitude zero points of 
26.581 and 28.087, respectively. These images are all aligned to the HFF HST 
images.
{Figure~\ref{fig:nircamimg} shows the NIRCam image cutouts around Y1, JD, and the new object f090d\_018.}

\subsection{NIRCam Wide-field Slitless Spectroscopy}

   MACS0416 was also observed by the NIRCam instrument in its wide-field 
slitless spectroscopy (WFSS) mode. This mode has two settings, Grism R and 
Grism C, which disperse light in either the direction of detector rows (``R'') 
or columns (``C'') in the LW channel, both having spectral resolution 
$R\approx 1600$ at $\sim$4~$\mu$m. The field was observed by the programs 
PID 2883 (``MAGNIF: Medium-band Astrophysics with the Grism of NIRCam in 
Frontier Fields,'' PI: F. Sun) and PID 3538 (``Unveiling the properties of 
high-redshift low/intermediate-mass galaxies in Lensing fields with NIRCam Wide 
Field Slitless Spectroscopy,'' PI: E. Iani). 

   The MAGNIF observations of MACS0416 were carried out on 2023 Aug 20 in 
Grism C using the F480M filter with total integration time of $\sim$3100 
seconds. These observations covered only JD and f090d\_018 with Y1
$\sim$3\arcsec\ outside the 
field. The observations of PID 3538 in this field were carried out on 2023 Dec 
22--24 and 2024 Jan 17 in both Grisms R and C\null. Four filters were used, 
F300M, F335M, F410M, and F460M with total integration times in each band 1761 
and 1546 seconds respectively for the two grisms. 

   To reduce the grism data, we retrieved the Level 1b ``uncal'' files
from the Mikulski Archive for Space Telescopes (MAST) and ran them through the
{\tt calwebb\_detector1} routine of the JWST data reduction pipeline (version 
1.13.4 in the context of jwst\_1223.pmap) to obtain the ``rate.fits'' files. 
We then followed the procedures described by \citet[][]{Sun2023_wfss} 
to further process the data. All single exposures were registered to the GAIA 
DR3 astrometry. There is a systematic offset between the HFF astrometry and that 
of GAIA in this field, which can be corrected by 
R.A.(GAIA) $=$ R.A.(HFF) $+$ 0\farcs21, and 
Decl.(GAIA) $=$ Decl.(HFF) $-$ 0\farcs08.
This was taken into account when extracting the spectra.

\section{Data Analysis} \label{sec:analysis}

\subsection{Overview}

\begin{figure*}[hbt!]
    \centering
    \includegraphics[width=\textwidth,height=\textheight,keepaspectratio]{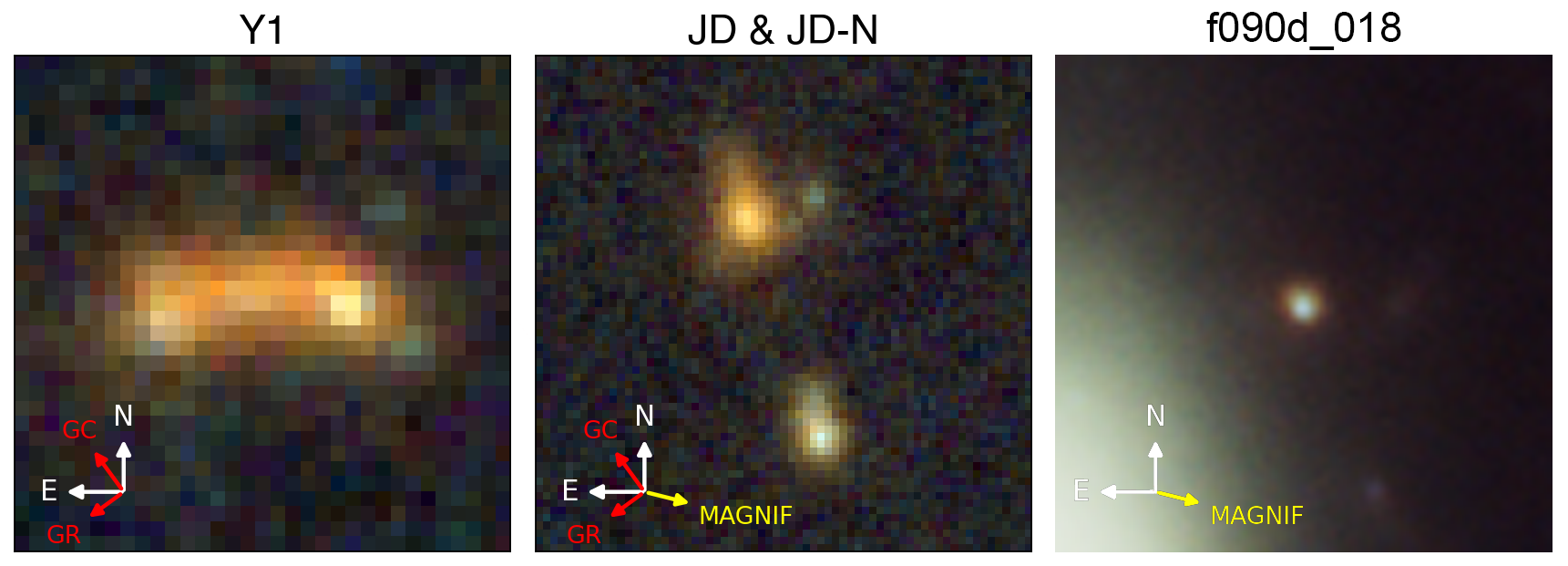}
    \caption{Color images of the three fields.  Image sizes are 1\arcsec\ squar for Y1 and 2\arcsec\ square for the other two fields. The color scheme is F090W+F115W+F150W as blue, 
    F200W+F277W as green, and F356W+F410M+F444W as red. In each image, the red 
    arrows show the NIRCam grism WFSS dispersion direction in the PID 3538 
    observation in Grism C (GC) and Grism R (GR), respectively; the yellow arrows in the other two images show
    the dispersion direction of GC utilized by the MAGNIF 
    observations. 
    }
    \label{fig:colorcomp}
\end{figure*}

    From their high-resolution (beam size 81.1$\times$112.2~mas) ALMA 
[\ion{O}{3}]~88~$\mu$m image of Y1, \citet{Tamura2023} resolved the 
system into three knots (``O1,'' ``O2,'' and ``O3''). Using this as a guide,
these authors separated its HST WFC3 image into three components (``E,'' ``C,'' 
and ``W'' from east to west) that coincide with the three 
[\ion{O}{3}]~88~$\mu$m knots. These three components are clearly 
distinguished in the NIRCam SW images (Figures~\ref{fig:nircamimg} 
and~\ref{fig:colorcomp}), and the entire system extends $\sim$0\farcs73 along 
the long axis.

   JD has a close companion object 0\farcs91 away to its northeast, which was
also selected by \citet{Laporte2015} as a Y-band dropout (their 
``MACSJ0416.1\_Y3''). As we will show below, its colors meet the requirements of
a $J$-band dropout as well, and its redshift is almost the same as that of JD. 
We refer to it as ``JD-N'' in this work to indicate its association with JD.

   {The object f090d\_018 is among a sample of F090W dropouts that we 
selected in this field (Sun et al., in prep.). While this object is very close 
to a bright member galaxy of the M0416 cluster 
\citep[at $z=0.401$; see][]{Kokorev2022}, we can still obtain reliable photometry for this object.
}

\subsection{Photometry and SEDs}

   To construct the SEDs of our targets, we carried out photometry on the 60\,mas images. 
{In addition to the NIRCam data, we also incorporated the HST WFC3 data
in F105W, F125W, F140W, and F160W obtained by the HFF program.\footnote{Available at \url{https://archive.stsci.edu/pub/hlsp/frontier/macs0416/images/hst/v1.0}.}
Although these WFC3 images are in the wavelength range covered by  NIRCam, their inclusion improves the wavelength sampling.
All the NIRCam and WFC3 images were convolved to match the point spread
function (PSF) of the F444W image, which has the coarsest spatial resolution (PSF full 
width at half-maximum  $\rm FWHM = 0\farcs145$). WebbPSFs 
\citep{Perrin2014, Perrin2015} were used in the convolution of the NIRCam images.
For the WFC3 images, we used the empirical PSFs available at the WFC3
instrumentation site.\footnote{\url{https://www.stsci.edu/hst/instrumentation/wfc3/data-analysis/psf}}
}
Matched-aperture photometry was done by running {\sc SExtractor} 
\citep{Bertin1996} in the dual-image mode with F444W as the detection band, and 
we adopted the \texttt{MAG\_ISO} magnitudes. 
{The magnitude uncertainties were calculated using root mean square (rms)
maps  derived with  {\sc astroRMS}.\footnote{Courtesy of M. Mechtley; see \url{https://github.com/mmechtley/astroRMS}} 
}

   {Two details affected the photometry: (1) JD-N has a 
faint neighbor that is not visible in F090W, but the neighbor does not pass our 
selection for F090W dropouts. The photometry for JD-N was obtained after masking 
this neighbor. (2) f090d\_018 could be affected by the outskirts of the 
$z=0.401$ foreground galaxy to the southeast, and therefore we modeled this 
foreground galaxy 
using  {\sc Galfit}  \citep[][]{Peng2002, Peng2010} and subtracted it 
from all images before doing photometry of f090d\_018.
}
   
\begin{figure*}
    \centering
    \includegraphics[width=\textwidth]{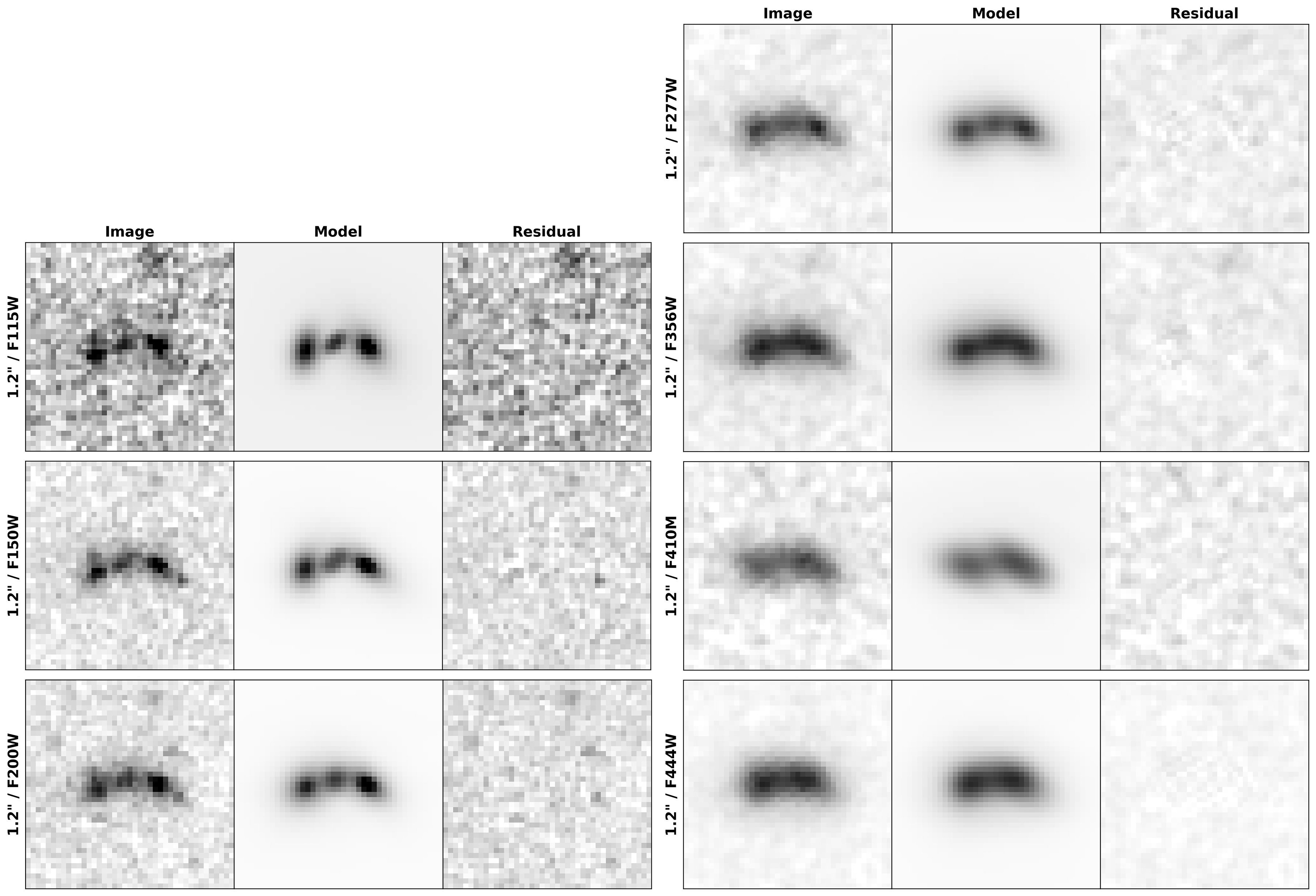} % "Y1_gafit" is the old one
    \caption{
    {Demonstration of the decomposition of Y1's three components in the
    seven NIRCam images as labeled. Each panel displays the original image 
    (left), the model constructed by {\sc GALFIT} using the S\'ersic profiles 
    (middle) and the residual image by subtracting the model from the original image (right). The images are all $1\farcs2\times 1\farcs2$ in size.}
    }
    \label{fig:y1decomp}
\end{figure*}

    {We also obtained the SEDs for Y1's three components by decomposing them
using {\sc Galfit} on the 30\,mas images (not convolved for the PSF sizes). Figure~\ref{fig:y1decomp} demonstrates the 
decomposition, which was done only for 
the NIRCam images but not the WFC3 ones because the latter do not 
have sufficient spatial resolution. Briefly, we used a single S\'ersic profile 
\citep[][]{Sersic1963} to model the light distribution of each component and fit 
the three components simultaneously. The centers of the three components were 
determined on the F200W image, where they are most clearly separated, and
the fit was done with the centers fixed. 
}
   Table~\ref{tbl:phot} summarizes the photometry for all sources. 
{For
Y1 as a whole and the other three objects, the 2$\sigma$ upper limits in F090W 
were calculated on the rms map at the source positions within a circular 
aperture whose size matches the {\tt MAG\_ISO} aperture or 
$\rm r=0\farcs2$, whichever is larger. For the three 
components of Y1, these upper limits were calculated within an $ r=0\farcs2$ 
circular aperture.
}
%From their colors (large magnitude breaks between F115W and F150W and being 
%${\rm S/N} <2$ in F090W), JD and JD-N indeed qualify as $J$-band dropouts.

\begin{table*}
    \centering
    \caption{NIRCam photometry (without correcting for lensing magnification)}
    \footnotesize
    \begin{tabular}{c|ccccccccc}
    \hline\hline
 & Y1 & Y1-E & Y1-C & Y1-W & JD & JD-N & f090d\_018 \\
 \hline
R.A. (deg) &  64.0391682 &  64.0392323 &  64.0391780 & 64.0391192  &  64.0479851 &  64.0480694 &  64.0428879 \\
Decl.\ (deg) & $-$24.0931764 & $-$24.0931976 & $-$24.0931800 & $-$24.0931886 & $-$24.0816685 & $-$24.0814279 & $-$24.0887917 \\
    \hline
R.A.$^s$ (deg) & 64.036045 &            &             &            &   64.0408500 &  64.0408680 &  64.038220 \\
Decl.$^s$ (deg) & $-$24.084671 &         &             &            & $-$24.0760100 & $-$24.0758710 & $-$24.0887910 \\
    \hline
F090W & $>$28.64  & $>$29.74 & $>$29.74 & $>$29.74 & $>$29.37 & $>$28.99 & $>$29.95 \\
F105W & $>$29.48 & ... & ... & ... & 28.63 $\pm$ 0.18 & 29.78 $\pm$ 0.71 & $>$30.01 \\ 
F115W & 26.94 $\pm$ 0.11 & 28.02 $\pm$ 0.07 & 29.09 $\pm$ 0.09 & 27.42 $\pm$ 0.15 & 27.68 $\pm$ 0.23 & 27.91 $\pm$ 0.20 & 28.27 $\pm$ 0.10 \\
F125W & 26.55 $\pm$ 0.06 & ... & ... & ... & 26.92 $\pm$ 0.05 & 27.38 $\pm$ 0.10 & 27.71 $\pm$ 0.07 \\ 
F140W & 26.24 $\pm$ 0.04 & ... & ... & ... & 26.84 $\pm$ 0.04 & 26.98 $\pm$ 0.07 & 27.24 $\pm$ 0.05 \\ 
F150W & 26.16 $\pm$ 0.06 & 27.07 $\pm$ 0.04 & 28.15 $\pm$ 0.04 & 26.77 $\pm$ 0.06 & 26.72 $\pm$ 0.05 & 26.69 $\pm$ 0.06 & 27.36 $\pm$ 0.05 \\
F160W & 26.12 $\pm$ 0.03 & ... & ... & ... & 26.87 $\pm$ 0.04 & 26.72 $\pm$ 0.05 & 27.17 $\pm$ 0.05 \\ 
F200W & 26.09 $\pm$ 0.05 & 26.77 $\pm$ 0.05 & 28.06 $\pm$ 0.08 & 26.90 $\pm$ 0.03 & 26.90 $\pm$ 0.05 & 26.59 $\pm$ 0.06 & 27.52 $\pm$ 0.05 \\
F277W & 25.98 $\pm$ 0.03 & 26.77 $\pm$ 0.03 & 27.69 $\pm$ 0.15 & 26.75 $\pm$ 0.03 & 27.05 $\pm$ 0.05 & 26.45 $\pm$ 0.03 & 27.87 $\pm$ 0.07 \\
F356W & 25.60 $\pm$ 0.02 & 26.13 $\pm$ 0.03 & 27.58 $\pm$ 0.19 & 26.50 $\pm$ 0.03 & 26.99 $\pm$ 0.04 & 26.06 $\pm$ 0.02 & 27.83 $\pm$ 0.06 \\
F410M & 25.92 $\pm$ 0.04 & 26.58 $\pm$ 0.03 & 27.58 $\pm$ 0.13 & 26.60 $\pm$ 0.03 & 27.13 $\pm$ 0.08 & 26.12 $\pm$ 0.04 & 27.85 $\pm$ 0.08 \\
F444W & 24.85 $\pm$ 0.01 & 25.92 $\pm$ 0.03 & 25.91 $\pm$ 0.05 & 25.94 $\pm$ 0.03 & 26.30 $\pm$ 0.03 & 25.42 $\pm$ 0.02 & 27.56 $\pm$ 0.03 \\
    \hline 
    \end{tabular}
    \raggedright
    \tablecomments{
{The R.A. and Decl.\ values in the first two rows are the observed
positions in the HFF frames, i.e., the offsets in Section~2 need to be applied to put them in the GAIA frame. The R.A.$^s$ and Decl$^s$ values in the next two 
rows are the positions in the source plane, again in HFF coordinates.
The listed magnitudes for Y1 (as a whole), JD, JD-N, and f090d\_018 are 
the \textsc{SExtractor} \texttt{MAG\_ISO} values obtained on the PSF-matched 
images, using the same aperture as defined in F444W\null. Those for the three 
components of Y1 are the magnitudes obtained by decomposition using 
\textsc{GALFIT} on the non-PSF-matched images. (See Section 3.2 and 
Figure~\ref{fig:y1decomp}.) No decomposition was possible for the four HST WFC3 bands 
(F105W, F125W, F140W, and F160W) because of their coarse spatial resolution.
}
}\label{tbl:phot}
\end{table*}

\subsection{Spectroscopic Identifications}
   
   Figure \ref{fig:colorcomp} shows the color composites of these two systems
with the WFSS dispersion directions indicated. 
The NIRCam WFSS mode is very suitable for the detection of emission lines,
which is the main focus here. To optimize the line detection, we subtracted a 
continuum estimated (following \citealt{Kashino2023}) by running a median filter 
$1\times 51$ pixels in size with a 9-pixel ``hole'' at the center along each row
or column.  

\begin{figure*}
    \centering
    \includegraphics[width=\textwidth]{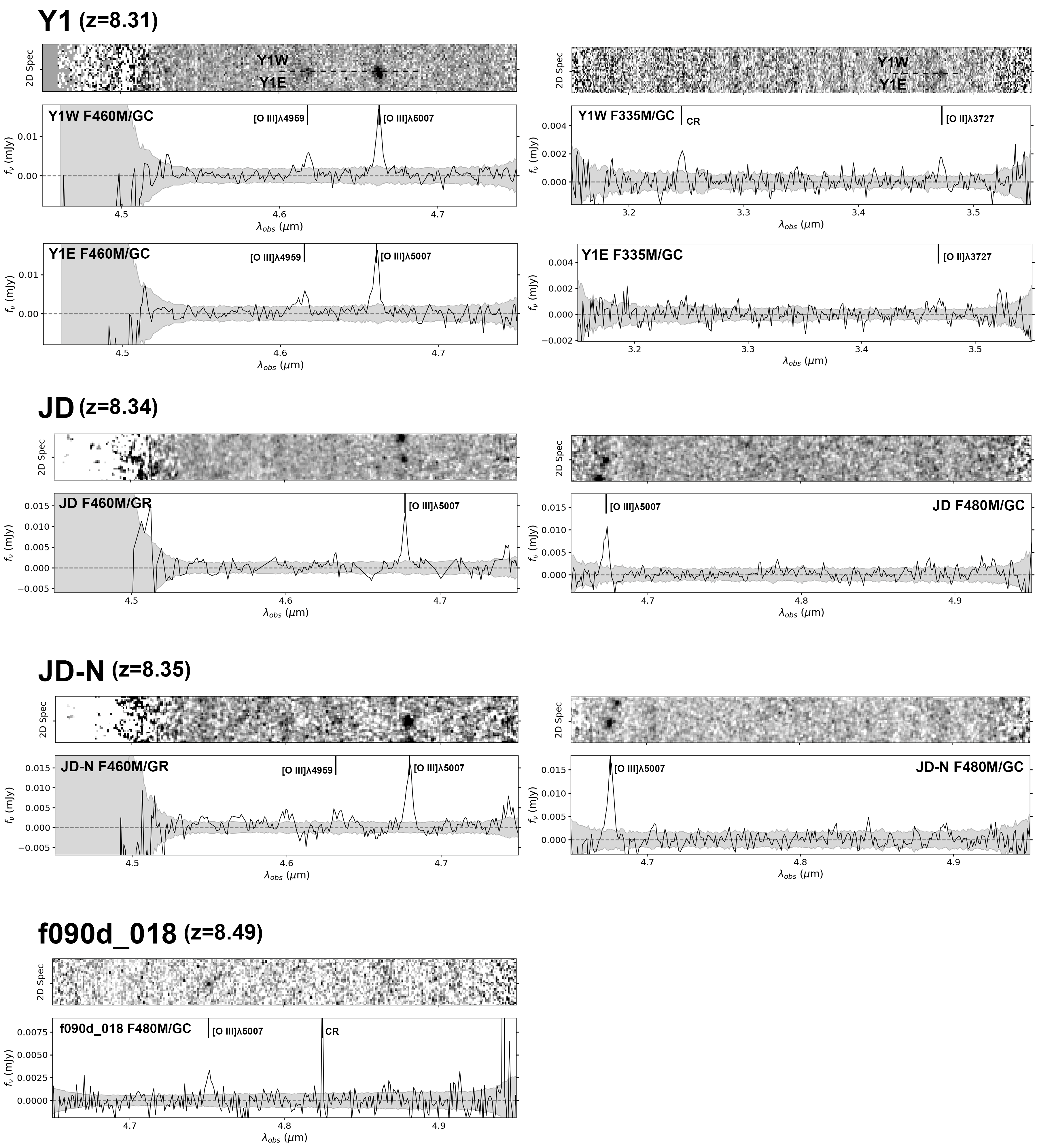}
    \caption{Extracted 2D and 1D spectra from the NIRCam WFSS observations.
    For each object, the top panel shows the 2D spectrum with background and 
    continuum subtracted, and the bottom panel shows the resultant 1D spectrum 
    with bandpass filter and grism orientation labeled. Identified emission lines are marked. The 2D 
    spectra of JD and JD-N are displayed after smoothing with a Gaussian kernel 
    with $\sigma=1$, while those of Y1 are not smoothed in order to show the 
    separation of the two components. 
    {The grey shading in the 1D spectra indicates the 1$\sigma$ uncertainty band.}
    With only a few exposures 
    in each setup, stacking has limited ability to remove outlying pixels, 
    leaving some narrow spikes in the 1D spectra. Fortunately, the emission 
    lines of interest are not contaminated by such spikes.
    }
    \label{fig:spec}
\end{figure*}

    As Y1 consists of three components, it would be ideal to extract the 
spectra for each component. In practice, however, the individual extraction 
could only split the signal into two parts, which largely coincide with Y1-E and 
Y1-W, respectively. This is due to both the faintness of Y1-C at 3--5~$\mu$m and 
the coarse resolution at these wavelengths. For simplicity, we refer to the 
separate extractions as Y1-E and Y1-W even though both could include some 
contribution from Y1-C\null. As shown in Figure \ref{fig:colorcomp}, the PID 
3538 Grism R dispersion direction is very close to the spatial extension of the 
Y1 system, causing severe spectral contamination between the two components. For
this reason,  we only used the Grism C data. The extracted 2D and 1D spectra are 
shown in Figure~\ref{fig:spec}. 

    For both Y1-E and Y1-W, the 
[\ion{O}{3}]\,$\lambda\lambda$4959,5007 lines are clearly 
seen in the F460M data. The detection of the 
[\ion{O}{2}]\,$\lambda$3727 line in the F335M data is marginal in 
the 1D spectrum but convincing in the 2D spectrum. To determine the 
redshifts, we fitted a Gaussian profile to the 
[\ion{O}{3}]\,$\lambda$5007 
line, which is the strongest, to obtain its observed central wavelengths. The 
redshifts thus derived are $z=8.309\pm0.002$ for Y1-E and $z=8.312\pm0.002$ for  
Y1-W\null.  We also fitted Gaussians to the weaker lines, using the initial guesses 
of the line centers at where they should be under these redshifts. The results 
based on the best-fit central wavelengths thus obtained are all in excellent 
agreement with the redshifts based on the 
[\ion{O}{3}]\,$\lambda$5007 line.

    For JD and JD-N, we extracted the spectra using the data from both MAGNIF 
and PID~3538. However, we discarded the Grism C data from the latter because 
its dispersion direction is almost parallel to the position angle of the two 
objects (Figure~\ref{fig:colorcomp}). The spectra are shown in 
Figure \ref{fig:spec}. JD-N shows two lines, one strong and the other marginal,  
in the F460M data from PID 3538 Grism R\null. Based on 
Gaussian fitting as above, these two lines coincide with the
[\ion{O}{3}]\,$\lambda\lambda$4959,5007 doublet at  
$z=8.346\pm0.002$. The MAGNIF F480M Grism C data reveal only one line, which
should be [\ion{O}{3}]\,$\lambda$5007. The 
[\ion{O}{3}]\,$\lambda$4959 line falls outside the filter 
transmission range and is not detected. For JD, the data from both programs show 
only one strong line. Given its possible redshift range, the line must be 
[\ion{O}{3}]\,$\lambda$5007 at $z=8.341\pm 0.002$. The 
non-detection of the [\ion{O}{3}]\,$\lambda$4959 line in the MAGNIF
data is again because it is outside of the filter transmission range. On the
other hand, the non-detection of the [\ion{O}{3}]\,$\lambda$4949 
line in the PID 3538 data is 
likely because of limited sensitivity: given the expected ratio of 
$\sim$1:3 of the [\ion{O}{3}] doublet, the 
[\ion{O}{3}]\,$\lambda$4959 line flux is comparable to the noise 
level of the data.

{The new object f090d\_018 was covered by both MAGNIF and PID 3538. However,
only the MAGNIF data were useful in determining a redshift.
For PID 3538, the GR  dispersion
direction pointed toward the nearby bright neighbor, which contaminates the data. No emission line was detected in the GC data. The MAGNIF
F480M GC data show one emission line at 4.752~$\mu$m and no
other obvious features.
The SED fitting puts this source at $z_{\rm ph}=8.75^{+0.04}_{-0.04}$, and we
attribute this line to [\ion{O}{3}]\,$\lambda$5007 because it is
the only reasonable choice (see Appendix~A). This gives 
$z=8.490\pm0.003$.
The [\ion{O}{3}]\,$\lambda$4959 line was not detected because of
its weakness. The non-detection of lines in PID 3538 can be explained by the
limited wavelength coverage:
the absence of the [\ion{O}{3}]\,$\lambda$5007 line in the F460M
data and the [\ion{O}{2}]\,$\lambda$3727 line in the F335M data are
all because the lines either fall in low-transmission region of the filter or is
completely out of the range.
}
 
    Table~\ref{tbl:nrc_wfss} summarizes the line measurements. The line widths
were converted to velocities by $\Delta v=c\Delta\lambda/\lambda_c$, where 
$\lambda_c$ and $\Delta\lambda$ are the mean and FWHM of the Gaussian fit, 
respectively, and $c$ is the speed of light. The total line intensities were 
obtained by integrating the fitted Gaussian profile within a $4\times$FWHM 
wavelength range centered at $\lambda_c$. The associated errors were estimated
using the non-smoothed 1D spectra. When calculating the observed 
equivalent widths (EW), the F410M magnitudes in Table~\ref{tbl:phot} were used 
as the continuum flux density. To account for continuum contributions from Y1-C 
in the spectra of Y1-E and Y1-W, half the F410M flux density of Y1-C was added 
to the Y1-E and Y1-W continua. Not surprisingly, these objects all have very 
large [\ion{O}{3}] EW values, which can largely account for the 
brightening of their SEDs in F444W\null. For example, the measured 
[\ion{O}{3}]\,$\lambda$5007 EW of JD implies that this line alone 
increases the object's F444W brightness (as compared to that in F410M) by 
$-0.83\pm0.06$ mag, which is almost the same as the observed 
$m_{410}-m_{444}=-0.83\pm0.09$~mag.

% in our adopted cosmology, Y1 has L_[OIII](4959+5007) = 6.1e9 L_sun
% Tamura et al. (2019) gives L_[OIII]88 = 1.2e9 L_sun
% the ratio is then 5.07. 

%\begin{table*}[hbt!]
\begin{table*}
    \caption{Summary of NIRCam WFSS results 
    }
    \centering
    \resizebox{\textwidth}{!}{
    \begin{tabular}{ccccccccccccc} \hline\hline
     Target & Program & Grating & $z$ & $I_{5007}$ & $\Delta v_{5007}$ & EW$_{5007}$ & $I_{4959}$ & $\Delta v_{4959}$ & EW$_{4959}$ & $I_{3727}$ & $\Delta v_{3727}$ & EW$_{3727}$ \\ 
        \hline
          Y1-E & 3538 & F335M+F460M/C & 8.309\textpm0.002 & 1.10\textpm0.12 & 338\textpm26 & 14879\textpm1221 & 0.43\textpm0.07 & 366\textpm91 & 5727\textpm733 & 0.18\textpm0.08 & 541\textpm171 & 1338\textpm542 \\ 
          Y1-W & 3538 & F335M+F460M/C & 8.312\textpm0.002 & 1.11\textpm0.10 & 310\textpm32 & 12519\textpm913 & 0.45\textpm0.11 & 346\textpm78 & 4967\textpm848 & 0.24\textpm0.10 & 424\textpm71 & 1505\textpm310 \\ 
        \hline
        JD & MAGNIF & F480M/C & 8.339\textpm0.002 & 0.55\textpm0.12 & 249\textpm35 & 7394\textpm1805 & ... & ... & ... & ... & ... & ... \\ 
          & 3538 & F335M+F460M/R & 8.341\textpm0.002 & 0.64\textpm0.10 & 217\textpm23 & 8635\textpm1124 & $<$3.86/5.39 & ... & ... & $<$2.06/5.11 & ... & ... \\ 
        \hline        
        JD-N & MAGNIF & F480M/C & 8.346\textpm0.002 & 1.13\textpm0.11 & 274\textpm35 & 5479\textpm423 & ... & ... & ... & ... & ... & ... \\         
          %& 3538 & F460M/R & 8.346\textpm0.002 & 1.07\textpm0.10 & 285\textpm20 & 5242\textpm504 & 0.32\textpm0.12 & 352\textpm98 & 1515\textpm588 & ... & ... & ... \\ 
          & 3538 & F335M+F460M/R & 8.346\textpm0.002 & 1.09\textpm0.12 & 316\textpm20 & 5325\textpm356 & 0.39\textpm0.15 & 409\textpm91 & 1870\textpm442 & $<$2.02/5.00 & ... & ... \\ 
        \hline 
        f090d\_018 & MAGNIF & F480M/C & 8.490\textpm0.003 & 0.18\textpm0.04 & 266\textpm46 & 5296\textpm1063 & ... & ... & ... & ... & ... & ... \\
        % & 3538 & F335M+F460M/C & ... & $<$11.08/14.63 & ... & ... & $<$5.04/6.83 & ... & ... & $<$6.08/14.46 \\
        \hline 
    \end{tabular}
    }
    \raggedright
    \tablecomments{The measurements were all obtained by fitting a Gaussian 
    profile to the emission lines (not corrected for the magnifications). 
    The spectroscopic redshifts are based on the fitted central wavelengths of 
    the [\ion{O}{3}]\,$\lambda$5007 line,
    which is the strongest.  The integrated line fluxes $I$ are in units of
    $10^{-17}$~erg~cm$^{-2}$~s$^{-1}$; the line widths $\Delta v$ were calculated 
    using the FWHM values of the lines and are in units of km~s$^{-1}$.
    The observed equivalent widths (EW, in \AA) were calculated assuming a flat 
    $f_{\nu}$ continuum at the level determined by the F410M magnitude. 
    The subscripts ``5007'', ``4959,'' and ``3727'' represent the
    [\ion{O}{3}]\,$\lambda$5007, 
    [\ion{O}{3}]\,$\lambda$4959, and 
    [\ion{O}{2}]\,$\lambda$3727 lines, respectively. 
    %For PID 3538, the [\ion{O}{3}] doublets were both identified in F460M Grism R, and the [\ion{O}{2}] line was identified in F335M Grism R. 
    {For the non-detections, the 2$\sigma$  
    %flux-density upper limits are given in the form of ``$< f_\nu/f_\lambda$'', 
    %where 
    $f_\nu$ and $f_\lambda$ upper limits are in units of $\mu$Jy and $10^{-20}$~erg~s$^{-1}$~cm$^{-2}$~\AA$^{-1}$, respectively. 
    }
    }
    \label{tbl:nrc_wfss}
\end{table*}

\section{Discussion} \label{sec:prop}

\subsection{Redshifts, magnification, and environment}

   The redshifts of Y1-E and Y1-W agree with each other within the 
uncertainties, and their rest-frame relative velocity is only 
$97\pm64$~km~s$^{-1}$. The average is $z=8.311\pm0.003$, which is in very good 
agreement with the previously reported $z=8.3118\pm0.0003$ based on 
[\ion{O}{3}]~88~$\mu$m 
\citep{Tamura2019} and $z=8.31132\pm0.00037$ based on 
[\ion{C}{2}]~157.5~$\mu$m \citep{Bakx2020} for Y1.
The apparent [\ion{O}{3}]~88~$\mu$m line flux measured by 
\citep{Tamura2019} is 0.66~Jy~km~s$^{-1}$, which corresponds to
$7.92\times 10^{-18}$~erg~cm$^{-2}$~s$^{-1}$ at the observed frequency
(364.377~GHz). Therefore, the line intensity ratio of 
[\ion{O}{3}]~$\lambda$5007 and 88~$\mu$m is 2.79.

    For JD, our redshift $z=8.341\pm0.002$ disagrees with the previously 
reported $z=9.28$ \citep{Laporte2021MNRAS}, which was based on the claimed
${\rm S/N\approx 6}$ detection of the [\ion{O}{3}]~88~$\mu$m line. 
That redshift could not explain the emission line observed at 4.68~\micron. 
{Therefore, we believe that their identification of the
[\ion{O}{3}]~88~$\mu$m line is incorrect and that the claimed
detection is likely due to a false positive. There have been a few cases in the
literature reporting seemingly significant millimeter line detections indicative
of high redshifts but being subsequently refuted 
\citep[e.g.,][gave recent examples]{Popping2023, Harikane2024}, and JD reinforces the need for caution.
}

    According to the lens model of \citet{Diego2023}, the magnification factors
for Y1, the JD/JD-N pair, and f090d\_018 are $\mu=1.21$, 2.26, and 1.32,
respectively. Based on their magnitudes in F150W (sampling the rest-frame UV 
range of $\sim$1430--1790\AA), their intrinsic absolute UV 
magnitudes after correcting for lensing magnifications are $M_{\rm UV} =-20.83$,
$-19.60$, $-19.77$, and $-19.57$~mag for Y1, JD, JD-N, and f090\_018, 
respectively. The observed size of Y1 as a whole is only marginally affected by 
the magnification, and its corresponding physical size 
along the long axis is $\sim$3.4~kpc. The separation of JD and JD-N, on the 
other hand, is affected significantly by lensing. Their separation in the 
source plane (see Table~\ref{tbl:phot}) is $\sim$0\farcs51, which corresponds to 
2.4~kpc. The velocity offset between JD and JD-N is only $160\pm64$~km~s$^{-1}$. 
Therefore, JD and JD-N should form an interacting pair.

   {After correcting for lensing magnification, the projected distance 
between Y1 and the JD/JD-N pair is
only $\sim$165~kpc, and  their co-moving radial distances differ
by only $\sim$8.2~Mpc. Therefore, these three objects are likely part of an
overdensity at $z\approx 8.3$. This is similar to the $z\approx 8.2$ overdensity
recently identified by \cite{Helton2023} in the GOODS-S field. Furthermore,
Y1+JD/JD-N might signpost an even large structure: f090d\_018 at $z=8.49$ is 
separated from Y1 by $\sim$77~kpc projected and $\sim$48.7~Mpc in the  radial 
direction, suggesting that it could be a member of a  filament-like structure 
stretching from the Y1+JD/JD-N complex. A future paper (Foo et al.,
in prep.)\ will present a detailed analysis of this possible structure.
}
%line-of-sight velocities are only different by
%$\sim$964~km~s$^{-1}$.

% z=8.31,  9026.9 Mpc
% z=8.34,  9035.1 Mpc (8.2 Mpc from Y1)
% z=8.49,  9075.6 Mpc (48.7 Mpc)
% separation between f090d_018 and Y1 in the source plane is 35.2", which 
\subsection{Stellar populations}\label{sec:bagpipes}

\begin{figure*}
    \centering
    \includegraphics[width=\textwidth,trim={0 1.6cm 0 0},clip]{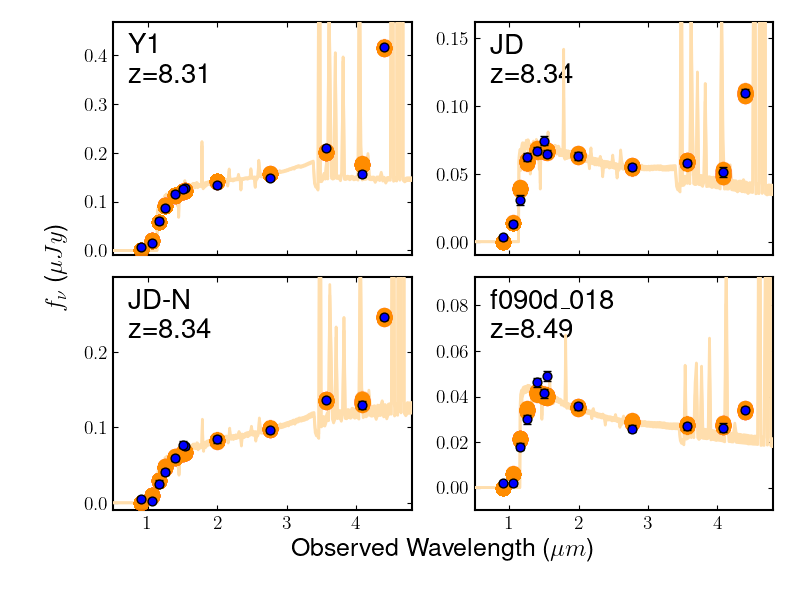} \\
    \includegraphics[width=\textwidth]{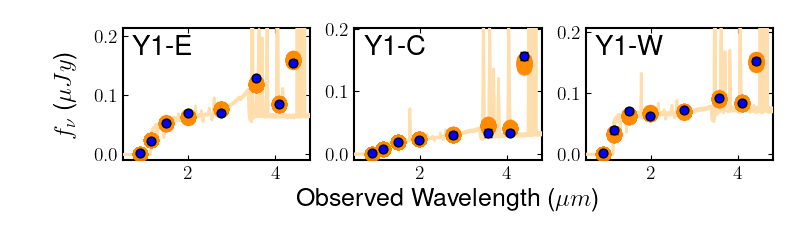}
    \caption{Illustration of the SED fitting with \textsc{Bagpipes} for each 
    entry in Table~\ref{tbl:phot}. The blue points with error bars show the 
    observed flux densities, and the orange points show the synthesized 
    magnitudes from the 50th-percentile fit. The posterior spectra 
    corresponding to the 16th and 84th percentile range are shown in yellow in 
    each panel.
    }
    \label{fig:sed_fitting}
\end{figure*}

\begin{figure*}
     \centering
    \includegraphics[width=0.4\textwidth]{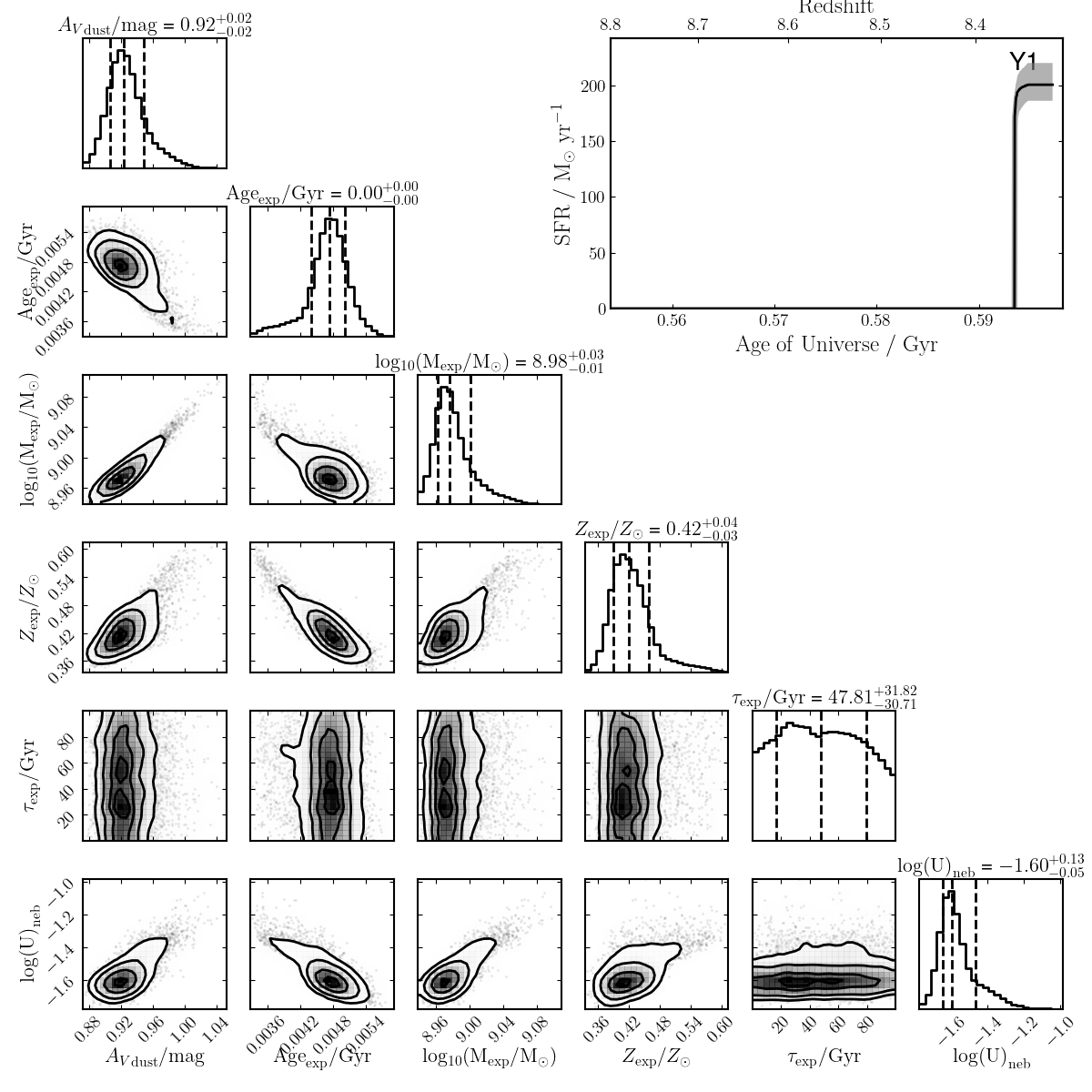}
    \includegraphics[width=0.4\textwidth]{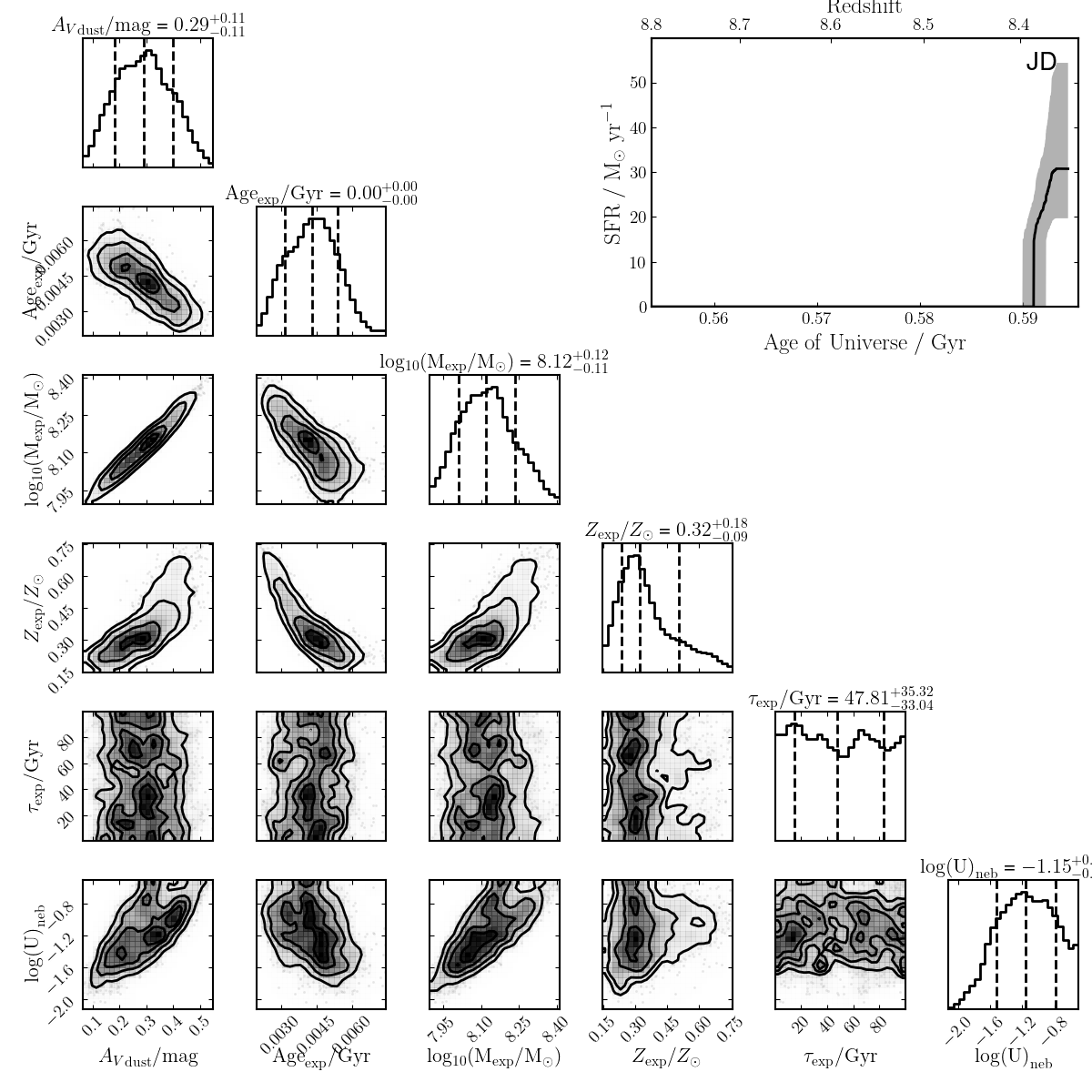} \\
    \includegraphics[width=0.4\textwidth]{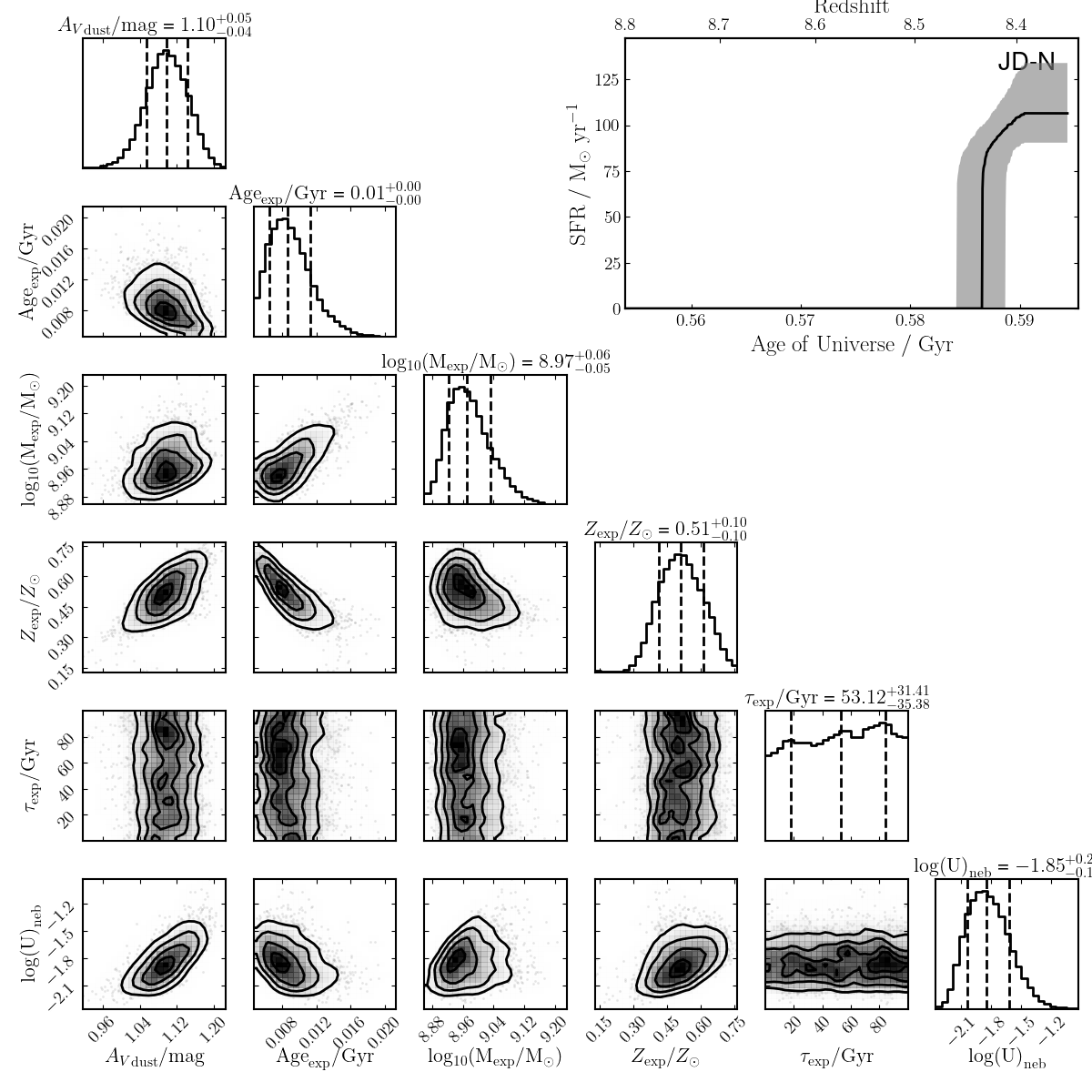} 
    \includegraphics[width=0.4\textwidth]{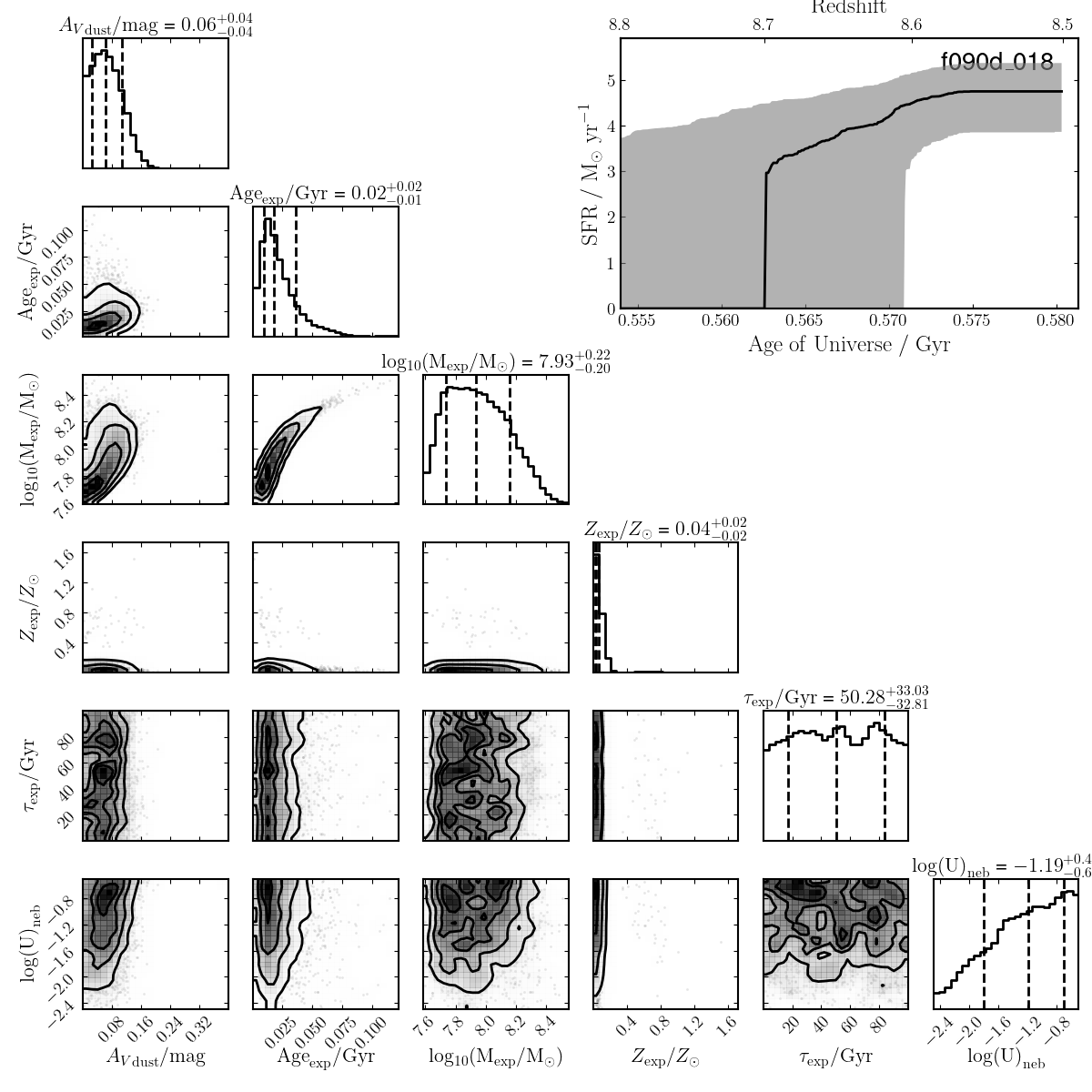}  \\
    \includegraphics[width=0.3\textwidth]{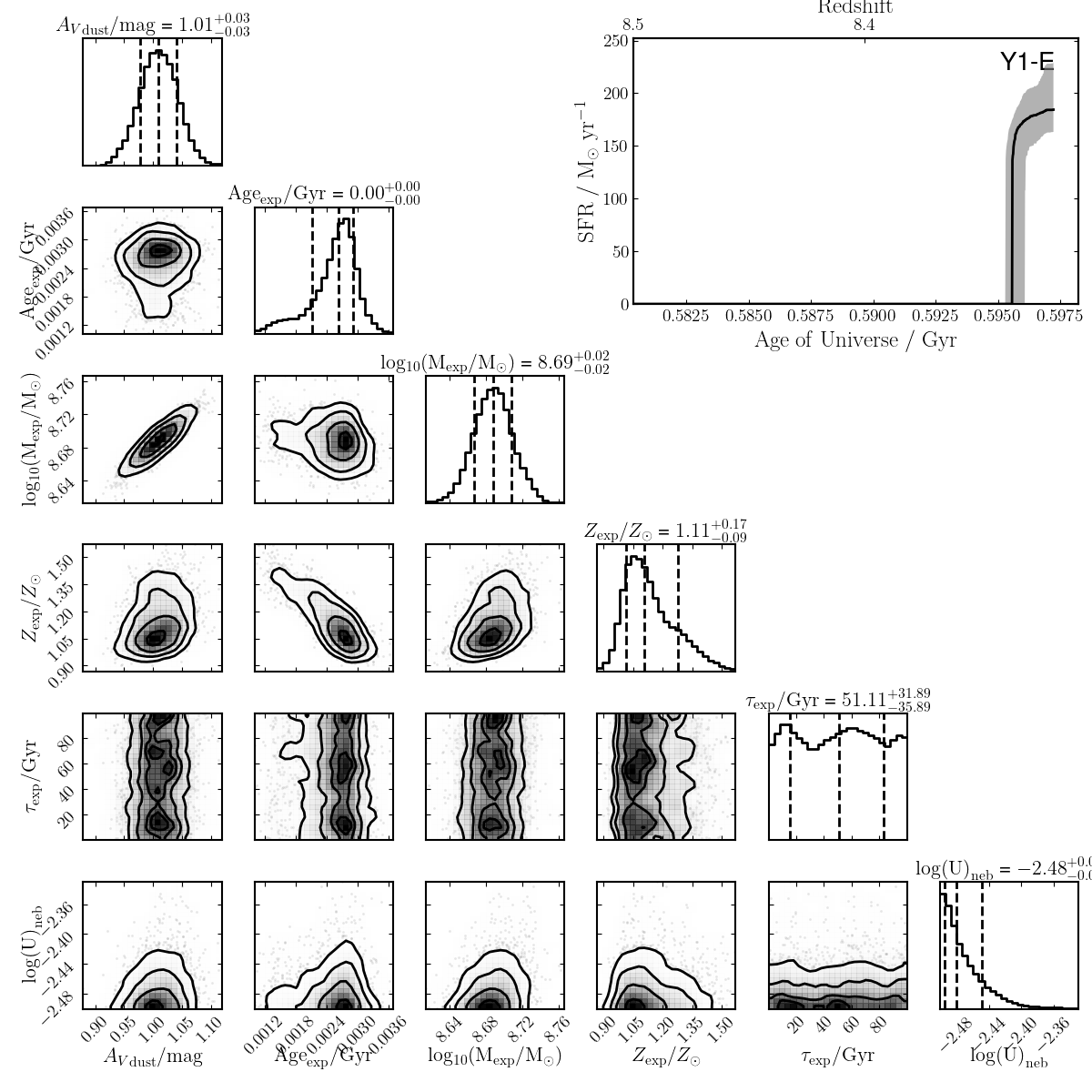}
    \includegraphics[width=0.3\textwidth]{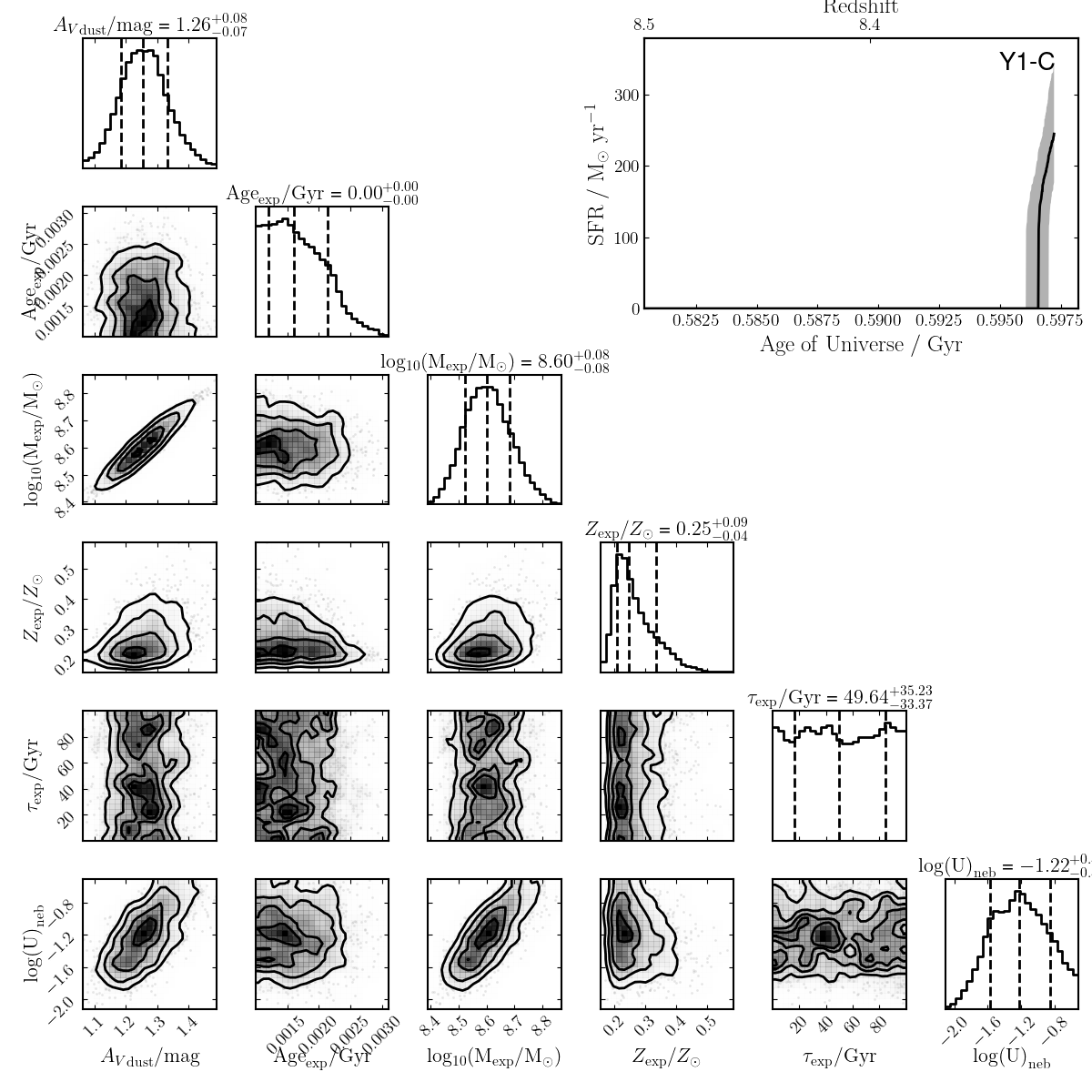}
    \includegraphics[width=0.3\textwidth]{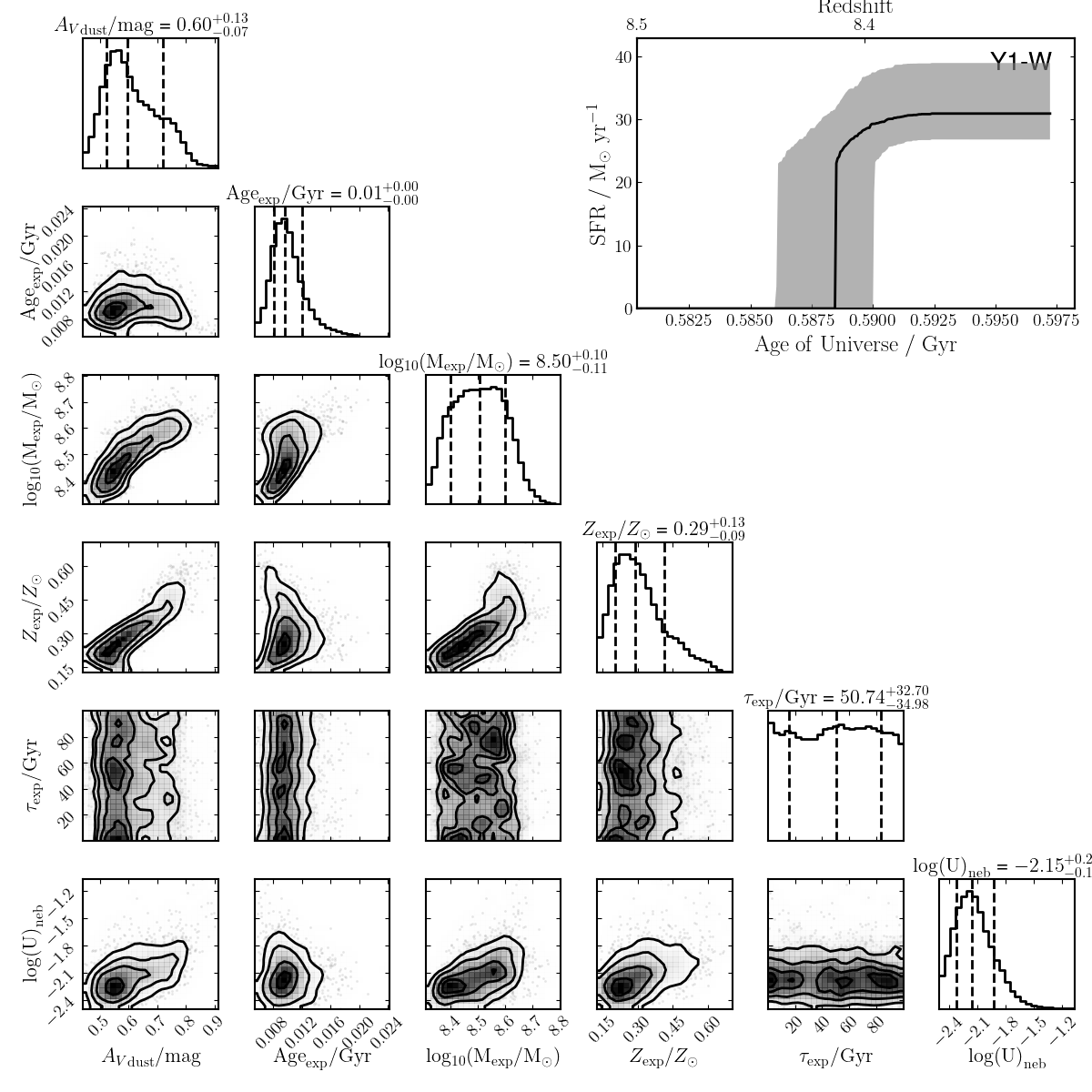}

    \caption{Corner plots showing the posterior distributions of the fitted 
    parameters corresponding to the SED fitting illustrated in 
    Figure~\ref{fig:sed_fitting}. The inset in the upper right corner in each
    panel shows the SFH, where the shaded region corresponds to the 16th to
    84th percentile range, and the black curve is the median value.}
    \label{fig:sed_fitting_corner}   
\end{figure*}

   To understand the stellar populations of these galaxies, we fitted their SEDs
using the \textsc{Bagpipes} software \citep{Carnall2018}, which utilizes the 
stellar population synthesis models of \citet{Bruzual2003} with the 
\citet{Kroupa2001} initial mass function. 
{The fitting was done at fixed redshifts of 8.31, 8.34 and 
8.49 for Y1, JD/JD-N, and f090d\_018, respectively.}
An exponentially declining star formation 
history (SFH) in the form of SFR~$\propto e^{-t/\tau}$ was assumed. The option
to include nebular emission lines was enabled, and we used the Calzetti 
dust-extinction law 
\citep{Calzetti1994, Calzetti2001} with $A_{V}$ ranging from 0 to 2.0~mag.
The metallicity was allowed in the range of $0 \le Z_{*}/Z_{\odot} \le 2.5$, and
the ionization parameter could vary in $-2.5 < {\rm log}(U) < -0.5$.
The results of the SED fitting are summarized in Table~\ref{tbl:sed_fitting}, and
Figure~\ref{fig:sed_fitting} shows the 16th to 84th percentile range of the 
posterior spectra superposed on the SEDs. The ``corner plots'' showing the 
posteriors for fitted parameters are given in 
Figure~\ref{fig:sed_fitting_corner}. 

{Overall, these four objects share many similarities, and the three 
components of Y1 are also quite similar to each other. In particular, Y1, JD 
and JD-N all have ages $\lesssim$10~Myr. To our 
knowledge, these are the youngest galaxy ages ever reported in the literature
\footnote{The $z=10.17$ galaxy reported in \cite{Hsiao2023} has a mass-weighted
age of 5--38~Myr depending on the SFH in use. We adopt the formation ages in 
this work, which are about twice as large as the mass-weighted ages for all our 
objects. 
}.
F090d\_018 is only about twice as old, albeit with a large uncertainty. The most 
distinct difference among them is in their SFRs, which span two orders of 
magnitude: after correcting the magnifications, the SFRs range from 3.6 (f090d\_018) to 200 (Y1)~${\rm M_\odot}$~yr$^{-1}$. 
}

Consistent with their young ages, the galaxies have low stellar masses of 
0.6--7.7~$\times 10^8~{\rm M_\odot}$ (after correcting for lensing 
magnifications). These can be achieved by the derived SFRs in the derived ages. (See 
also Figure~\ref{fig:sed_fitting_corner}.)
{Interestingly, f090d\_018, which has the lowest intrinsic stellar mass 
($5.7\times 10^7 {\rm M_\odot}$) and SFR (3.6~${\rm M_\odot}$~yr$^{-1}$), 
also has the smallest dust extinction, $A_{\rm V}\approx 0.06$~mag.
}

   The existence of the oxygen lines in these objects (as well as the
[\ion{C}{2}]~157.7 $\mu$m line in Y1) indicates that they have 
already acquired substantial amounts of metals. 
{This is supported by the metallicities derived from the SED 
analysis. Given the young 
ages for Y1, JD, and JD-N ($Z/Z_\odot\approx 0.3$--0.4), the only route for their metal enrichment was through core-collapse 
supernovae. On the other hand, the metallicity estimate for f090d\_018  ($Z/Z_\odot = 0.04^{+0.03}_{-0.02}$) is rather
low, which appears consistent with its
low SFR. 
}

\begin{table*}
\centering
\caption{\textsc{Bagpipes} posterior parameters of the fitted stellar populations}
\begin{tabular}{lcccccc}
\hline\hline
 Object & $A_{V}$ & Age/Myr & $Z/Z_{\sun}$ & ${\rm log}(U)$ & ${\rm log}(\mu M/\mathrm{M_{\sun}})$ & $\mu$SFR/($\mathrm{M_\odot\ yr}^{-1}$) \\
\hline

\hline\hline
Y1 & $0.92^{+0.03}_{-0.02}$ & $4.76^{+0.28}_{-0.35}$ & $0.42^{+0.04}_{-0.03}$ & $-1.60^{+0.14}_{-0.05}$ & $8.97^{+0.03}_{-0.01}$ & $200.05^{+22.66}_{-15.88}$ \\
JD & $0.29^{+0.11}_{-0.11}$ & $4.28^{+1.06}_{-1.17}$ & $0.32^{+0.19}_{-0.08}$ & $-1.15^{+0.35}_{-0.37}$ & $8.12^{+0.12}_{-0.11}$ & $30.39^{+23.98}_{-10.10}$ \\
JD-N & $1.10^{+0.05}_{-0.04}$ & $8.69^{+2.46}_{-2.19}$ & $0.51^{+0.10}_{-0.09}$ & $-1.85^{+0.23}_{-0.17}$ & $8.95^{+0.05}_{-0.04}$ & $108.25^{+26.44}_{-17.55}$ \\
f090d\_018 & $0.06^{+0.04}_{-0.04}$ & $17.79^{+15.50}_{-7.56}$ & $0.04^{+0.03}_{-0.02}$ & $-1.23^{+0.49}_{-0.58}$ & $7.88^{+0.19}_{-0.17}$ & $4.72^{+0.65}_{-0.80}$ \\
\hline
Y1-E & $1.01^{+0.03}_{-0.04}$ & $2.62^{+0.27}_{-0.48}$ & $1.11^{+0.18}_{-0.09}$ & $-2.48^{+0.03}_{-0.01}$ & $8.69^{+0.02}_{-0.02}$ & $186.01^{+61.84}_{-19.98}$ \\
Y1-C & $1.25^{+0.07}_{-0.07}$ & $1.62^{+0.52}_{-0.43}$ & $0.24^{+0.08}_{-0.04}$ & $-1.22^{+0.36}_{-0.35}$ & $8.60^{+0.08}_{-0.08}$ & $246.38^{+103.25}_{-67.36}$ \\
Y1-W & $0.60^{+0.13}_{-0.07}$ & $9.67^{+2.12}_{-1.65}$ & $0.29^{+0.14}_{-0.09}$ & $-2.16^{+0.24}_{-0.16}$ & $8.48^{+0.10}_{-0.10}$ & $30.86^{+8.17}_{-4.03}$ \\
\hline
\end{tabular}
\raggedright
\tablecomments{The quoted value for each parameter is the median of the 
posterior distribution, while the lower and upper error bars correspond to the 
16th and 84th percentile values. The SFRs are the instantaneous values at the 
latest age bins, which have width 5773 years. These values, as well as the 
stellar masses, are not corrected for the lensing magnifications, which 
(based on the \citealt{Diego2023} lens model) are $\mu=1.21$ for Y1, $\mu=2.26$ 
for JD and JD-N, and $\mu=1.32$ for f090d\_018.
}\label{tbl:sed_fitting}
\end{table*}

\begin{table*}[ht]
    \caption{Rest-frame UV slopes and $f_{\rm esc}$}
    \label{tbl:uv_slope}
    \centering
\begin{tabular}{lcccccc}
    \hline\hline
Name & $\beta^{150,200}$ & $f_{\rm esc}^{150,200}$ & $\beta^{150,277}$ & $f_{\rm esc}^{150,277}$ & $\beta^{\rm PL}$ & $f_{\rm esc}^{\rm PL}$ \\
\hline
% original version
% Y1   & $-$1.769 & 0.019 & $-$1.731 & 0.017 & $-$1.417 & 0.007 \\
% Y1-E & $-$1.838 & 0.023 & $-$1.955 & 0.032 & $-$1.304 & 0.005 \\
% Y1-C & $-$1.515 & 0.009 & $-$1.610 & 0.012 & $-$1.200 & 0.004 \\
% Y1-W & $-$1.761 & 0.018 & $-$1.834 & 0.022 & $-$1.206 & 0.004 \\
% \hline
% JD   & $-$2.630 & 0.210 & $-$2.479 & 0.138 & $-$1.967 & 0.033 \\
% JD-N & $-$1.718 & 0.016 & $-$1.673 & 0.014 & $-$1.204 & 0.004 \\
% rev1 20240814
Y1         & $-$1.775 & 0.019 & $-$1.730 & 0.017 & $-$1.165 & 0.003 \\
JD         & $-$2.581 & 0.183 & $-$2.486 & 0.140 & $-$2.168 & 0.057 \\
JD-N       & $-$1.691 & 0.015 & $-$1.639 & 0.013 & $-$0.952 & 0.002 \\
f090d\_018 & $-$2.522 & 0.155 & $-$2.772 & 0.313 & $-$2.404 & 0.111 \\
\hline
Y1-E       & $-$1.052 & 0.002 & $-$1.560 & 0.010 & $-$1.010 & 0.002 \\
Y1-C       & $-$1.688 & 0.015 & $-$1.295 & 0.005 & $-$0.845 & 0.001 \\
Y1-W       & $-$2.425 & 0.118 & $-$1.968 & 0.033 & $-$1.639 & 0.013 \\
\hline
\end{tabular}
\raggedright
\tablecomments{The UV slope $\beta$ is defined via
$f_\nu\propto \lambda^{\beta+2}$. The numerical superscripts on $f_{\rm esc}$  
indicate the passband pairs (F150W, F200W) and (F150W, F277W), respectively, 
while ``PL'' indicates the power-law fit to the best-fit template spectrum.
}
\end{table*}

\subsection{UV slopes and Lyman-continuum photon escape fractions}

  Using the photometry reported in Table~\ref{tbl:phot}, we calculated the 
rest-frame UV slope $\beta$ for these sources. The calculation was done using 
two different filter pairs, (F150W, F200W) and (F150W, F277W), which
approximately correspond to using the rest-frame wavelength pairs of 
(1600\AA, 2100\AA) and
(1600\AA, 3000\AA), respectively. As an alternative, we also used the best-fit 
template spectrum (corresponding to the 50th percentile of the posterior 
distribution) of the SED and fitted a power law between 1300\AA\ and 1850\AA\ in
the rest-frame to obtain this slope. These results are listed in
Table~\ref{tbl:uv_slope}. The slopes derived using the best-fit template spectra 
($\beta^{\rm PL}$) are all significantly flatter than those obtained using the 
photometry in the blue passband pairs ($\beta^{150,200}$ and $\beta^{150,277}$), 
which cautions that care must be taken when interpreting 
UV slopes obtained by different methods \citep[e.g.,][]{Dunlop2013, Austin2024}. 

   The UV slopes are related to the Lyman continuum photon escape fractions
($f_{\rm esc}$), which are critical in understanding the sources of the cosmic 
hydrogen reionization. For each $\beta$ value, we calculated $f_{\rm esc}$ 
following Equation~11 of 
\citet{Chisholm2022}, 
\begin{align}
    f_{\rm esc} = (1.3\pm 0.6) \times 10^{-4} \cdot 10^{(-1.22\pm 0.1) \beta} \,.
\end{align}
The derived values are reported in Table~\ref{tbl:uv_slope}.

{Somewhat surprisingly, only JD and f090d\_018 have blue UV slopes 
$\beta\lesssim -2.0$.  These two objects' inferred $f_{\rm esc}$ values are between $\sim$6\% and 
31\% depending on the adopted $\beta$. We believe that this can be
attributed to their much less dust extinctions ($A_{\rm V}\sim 0.3$ and 0.06~mag)
as compared to those of Y1 and JD-N ($A_{\rm V}\sim 0.9$ and 1.1~mag).
}
The existence of two objects with high extinctions and low escape fractions suggests 
that a significant fraction of very young galaxies at high redshifts might not 
contribute to reionization because they formed a considerable amount of dust that 
dispersed throughout their bodies in only a few Myr and blocks UV emissions.

\section{Summary}

{New JWST data have identified four very young galaxies at 
$z=8.31$--8.49 behind the lensing cluster MACS0416.}
Three of them are strong [\ion{O}{3}]\,$\lambda$5007 emitters, and
their line equivalent widths can easily explain their apparent flux excess in 
F444W\null. Two galaxies, Y1 and JD, had prior spectroscopic redshifts from 
ALMA detections of the [\ion{O}{3}]~88~$\mu$m line and/or 
[\ion{C}{2}]~157.7~$\mu$m line. We confirm the redshift 
8.31 for Y1 but find $z=8.34$ for JD as opposed to the previously claimed 
$z=9.28$. Y1 is a merging system that is resolved into three components, 
extending $\sim$3.4~kpc along the long axis. 
%The three components all have very similar properties. 
Another object, JD-N, is a previously discovered $z > 8$ candidate 
that is now confirmed to have the same redshift as JD; the two 
are only $\sim$0\farcs51 ($\sim$2.4~kpc) apart in the source plane and 
therefore very likely merging or at least interacting. 
{The final object, f090d\_018 at $z=8.49$, is newly identified. These four 
objects spread over a projected distance of
$\sim$165~kpc in the source plane and a radial distance of $\sim$49~Mpc,
and it is likely that they are part of an overdensity on a filament-like 
structure.
}

The four objects are magnified by less than a magnitude and have intrinsic 
$M_{\rm UV}$ ranging from $-19.57$ to $-20.83$~mag. Our SED analysis show that 
they are all very young systems in the making, with ages of $\sim$4--18~Myr,
arguably the youngest galaxies ever reported at $z>6$. These infant galaxies
have stellar masses on the order of $10^{7-8}$~${\rm M_\odot}$, and their SFRs 
range from a few to over a hundred $\rm{M_\odot}$~yr$^{-1}$. However, only JD
and f090d\_018 have blue rest-frame UV slopes ($\beta$ ranging from $-2.2$ to 
$-2.8$ depending on how it is derived) indicative of high Lyman-continuum 
photon escape fractions ($f_{\rm esc}$ possibly as high as $\sim$31\%). This is 
largely due to their small dust extinction ($A_{\rm V}=0.06$--0.29~mag). 
{Interestingly, these two objects are the least massive
($M_*=5.7$ and $5.8\times 10^7 {\rm M_\odot}$) and least active 
(${\rm SFR}=3.6$ and 13.6~${\rm M_\odot}$~yr$^{-1}$) ones among the four.
}
In contrast, Y1 and JD-N, despite their much higher SFRs, have 
$\beta \gtrsim -1.8$ to $-1.0$
(implying $f_{\rm esc}\lesssim 2\%$)  because of their higher dust extinction
($A_{\rm V}=0.92$--1.10~mag). This
suggests that even very young, very actively star-forming galaxies at high-$z$ 
could have negligible contribution to the ionizing background if they form dust 
throughout their bodies too quickly (over a few Myr time scale). Y1 and JD-N
are examples that dust formation and pollution processes at $z>8$ could indeed 
be very fast.\\

The NIRCam imaging and WFSS data used in this article were obtained from the 
Mikulski Archive for Space Telescopes (MAST) at the Space Telescope Science Institute. The specific observations analyzed can be accessed via \dataset[doi: 10.17909/6qc6-4g31]{https://doi.org/10.17909/6qc6-4g31}. 

\begin{acknowledgments}

We thank the anonymous referee for the constructive comments that have improved 
the quality of this paper.
We dedicate this work to the memory of our dear colleague Mario Nonino, a
kind and gentle person and an example for many. This project is based on 
observations made with the NASA/ESA/CSA James Webb
Space Telescope and obtained from the Mikulski Archive for Space Telescopes,
which is a collaboration between the Space Telescope Science Institute 
(STScI/NASA), the Space Telescope European Coordinating Facility (ST-ECF/ESA),
and the Canadian Astronomy Data Centre (CADC/NRC/CSA). We thank the Program 
Coordinator, Tony Roman, for his expert help scheduling the PEARLS program.
The authors also acknowledge the JWST GO-3538 team led by PI E. Iani for 
developing their observing program with a zero-exclusive-access period.
\end{acknowledgments}
ZM is supported by the NSF grants No. 1636621, 2034318, and 2307448.
CC is supported by the National Natural Science Foundation of China, 
No. 11933003 and 12173045 as well as the CAS through a grant to the CAS South 
America Center for Astronomy (CASSACA). 
BS and HY acknowledge the support from the NSF grant AST-2307447 and the University
of Missouri Research Council grant URC-23-029.
SHC, RAW, and RAJ acknowledge support from NASA JWST Interdisciplinary Scientist 
grants NAG5-12460, NNX14AN10G, and 80NSSC18K0200 from GSFC.
FS and CNAW acknowledges funding from the JWST/NIRCam contract NASS-0215 to the
University of Arizona.
FS acknowledges support for program \#2883 provided by NASA through a grant from 
the Space Telescope Science Institute, which is operated by the Association of 
Universities for Research in Astronomy, Inc., under NASA contract NAS 5-03127.
DE acknowledges support from a Beatriz Galindo senior fellowship (BG20/00224) 
from the Spanish Ministry of Science and Innovation, projects
PID2020-114414GB-100 
and PID2020-113689GB-I00 financed by MCIN/AEI/10.13039/501100011033, project 
P20-00334 financed by the Junta de Andaluc\'{i}a, and project A-FQM-510-UGR20 of 
the FEDER/Junta de Andaluc\'{i}a-Consejer\'{i}a de Transformaci\'{o}n Econ\'{o}mica, Industria, Consejer\'a, Universidades.

%\bibliography{refs}{}

\begin{thebibliography}{}
\expandafter\ifx\csname natexlab\endcsname\relax\def\natexlab#1{#1}\fi
\providecommand{\url}[1]{\href{#1}{#1}}
\providecommand{\dodoi}[1]{doi:~\href{http://doi.org/#1}{\nolinkurl{#1}}}
\providecommand{\doeprint}[1]{\href{http://ascl.net/#1}{\nolinkurl{http://ascl.net/#1}}}
\providecommand{\doarXiv}[1]{\href{https://arxiv.org/abs/#1}{\nolinkurl{https://arxiv.org/abs/#1}}}

\bibitem[{{Atek} {et~al.}(2024){Atek}, {Labb{\'e}}, {Furtak}, {Chemerynska},
  {Fujimoto}, {Setton}, {Miller}, {Oesch}, {Bezanson}, {Price}, {Dayal},
  {Zitrin}, {Kokorev}, {Weaver}, {Brammer}, {Dokkum}, {Williams}, {Cutler},
  {Feldmann}, {Fudamoto}, {Greene}, {Leja}, {Maseda}, {Muzzin}, {Pan},
  {Papovich}, {Nelson}, {Nanayakkara}, {Stark}, {Stefanon}, {Suess}, {Wang}, \&
  {Whitaker}}]{Atek2024}
{Atek}, H., {Labb{\'e}}, I., {Furtak}, L.~J., {et~al.} 2024, \nat, 626, 975,
  \dodoi{10.1038/s41586-024-07043-6}

\bibitem[{{Austin} {et~al.}(2024){Austin}, {Conselice}, {Adams}, {Harvey},
  {Duan}, {Trussler}, {Li}, {Juodzbalis}, {Ormerod}, {Ferreira}, {Westcott},
  {Harris}, {Wilkins}, {Bhatawdekar}, {Caruana}, {Coe}, {Cohen}, {Driver},
  {D'Silva}, {Frye}, {Furtak}, {Grogin}, {Hathi}, {Holwerda}, {Jansen},
  {Koekemoer}, {Marshall}, {Nonino}, {Ortiz}, {Pirzkal}, {Robotham}, {Ryan},
  {Summers}, {Willmer}, {Windhorst}, {Yan}, \& {Zackrisson}}]{Austin2024}
{Austin}, D., {Conselice}, C.~J., {Adams}, N.~J., {et~al.} 2024, arXiv
  e-prints, arXiv:2404.10751, \dodoi{10.48550/arXiv.2404.10751}

\bibitem[{{Bakx} {et~al.}(2020){Bakx}, {Tamura}, {Hashimoto}, {Inoue}, {Lee},
  {Mawatari}, {Ota}, {Umehata}, {Zackrisson}, {Hatsukade}, {Kohno}, {Matsuda},
  {Matsuo}, {Okamoto}, {Shibuya}, {Shimizu}, {Taniguchi}, \&
  {Yoshida}}]{Bakx2020}
{Bakx}, T. J.~L.~C., {Tamura}, Y., {Hashimoto}, T., {et~al.} 2020, \mnras, 493,
  4294, \dodoi{10.1093/mnras/staa509}

\bibitem[{{Bertin} \& {Arnouts}(1996)}]{Bertin1996}
{Bertin}, E., \& {Arnouts}, S. 1996, \aaps, 117, 393,
  \dodoi{10.1051/aas:1996164}

\bibitem[{{Bouwens} {et~al.}(2006){Bouwens}, {Illingworth}, {Blakeslee}, \&
  {Franx}}]{Bouwens2006}
{Bouwens}, R.~J., {Illingworth}, G.~D., {Blakeslee}, J.~P., \& {Franx}, M.
  2006, \apj, 653, 53, \dodoi{10.1086/498733}

\bibitem[{{Bouwens} {et~al.}(2010){Bouwens}, {Illingworth}, {Oesch}, {Trenti},
  {Stiavelli}, {Carollo}, {Franx}, {van Dokkum}, {Labb{\'e}}, \&
  {Magee}}]{Bouwens2010}
{Bouwens}, R.~J., {Illingworth}, G.~D., {Oesch}, P.~A., {et~al.} 2010, \apjl,
  708, L69, \dodoi{10.1088/2041-8205/708/2/L69}

\bibitem[{{Bruzual} \& {Charlot}(2003)}]{Bruzual2003}
{Bruzual}, G., \& {Charlot}, S. 2003, \mnras, 344, 1000,
  \dodoi{10.1046/j.1365-8711.2003.06897.x}

\bibitem[{{Calzetti}(2001)}]{Calzetti2001}
{Calzetti}, D. 2001, \pasp, 113, 1449, \dodoi{10.1086/324269}

\bibitem[{{Calzetti} {et~al.}(1994){Calzetti}, {Kinney}, \&
  {Storchi-Bergmann}}]{Calzetti1994}
{Calzetti}, D., {Kinney}, A.~L., \& {Storchi-Bergmann}, T. 1994, \apj, 429,
  582, \dodoi{10.1086/174346}

\bibitem[{{Carnall} {et~al.}(2018){Carnall}, {McLure}, {Dunlop}, \&
  {Dav{\'e}}}]{Carnall2018}
{Carnall}, A.~C., {McLure}, R.~J., {Dunlop}, J.~S., \& {Dav{\'e}}, R. 2018,
  \mnras, 480, 4379, \dodoi{10.1093/mnras/sty2169}

\bibitem[{{Chisholm} {et~al.}(2022){Chisholm}, {Saldana-Lopez}, {Flury},
  {Schaerer}, {Jaskot}, {Amor{\'\i}n}, {Atek}, {Finkelstein}, {Fleming},
  {Ferguson}, {Fern{\'a}ndez}, {Giavalisco}, {Hayes}, {Heckman}, {Henry}, {Ji},
  {Marques-Chaves}, {Mauerhofer}, {McCandliss}, {Oey}, {{\"O}stlin},
  {Rutkowski}, {Scarlata}, {Thuan}, {Trebitsch}, {Wang}, {Worseck}, \&
  {Xu}}]{Chisholm2022}
{Chisholm}, J., {Saldana-Lopez}, A., {Flury}, S., {et~al.} 2022, \mnras, 517,
  5104, \dodoi{10.1093/mnras/stac2874}

\bibitem[{{Citro} {et~al.}(2024){Citro}, {Scarlata}, {Mantha}, {Williams},
  {Rafelski}, {Revalski}, {Hayes}, {Henry}, {Rutkowski}, \&
  {Teplitz}}]{Citro2024}
{Citro}, A., {Scarlata}, C.~M., {Mantha}, K.~B., {et~al.} 2024, arXiv e-prints,
  arXiv:2406.07618, \dodoi{10.48550/arXiv.2406.07618}

\bibitem[{{Coe} {et~al.}(2015){Coe}, {Bradley}, \& {Zitrin}}]{Coe2015}
{Coe}, D., {Bradley}, L., \& {Zitrin}, A. 2015, \apj, 800, 84,
  \dodoi{10.1088/0004-637X/800/2/84}

\bibitem[{{Coe} {et~al.}(2019){Coe}, {Salmon}, {Brada{\v{c}}}, {Bradley},
  {Sharon}, {Zitrin}, {Acebron}, {Cerny}, {Cibirka}, {Strait},
  {Paterno-Mahler}, {Mahler}, {Avila}, {Ogaz}, {Huang}, {Pelliccia}, {Stark},
  {Mainali}, {Oesch}, {Trenti}, {Carrasco}, {Dawson}, {Rodney}, {Strolger},
  {Riess}, {Jones}, {Frye}, {Czakon}, {Umetsu}, {Vulcani}, {Graur}, {Jha},
  {Graham}, {Molino}, {Nonino}, {Hjorth}, {Selsing}, {Christensen},
  {Kikuchihara}, {Ouchi}, {Oguri}, {Welch}, {Lemaux}, {Andrade-Santos}, {Hoag},
  {Johnson}, {Peterson}, {Past}, {Fox}, {Agulli}, {Livermore}, {Ryan}, {Lam},
  {Sendra-Server}, {Toft}, {Lovisari}, \& {Su}}]{Coe2019}
{Coe}, D., {Salmon}, B., {Brada{\v{c}}}, M., {et~al.} 2019, \apj, 884, 85,
  \dodoi{10.3847/1538-4357/ab412b}

\bibitem[{{Cullen} {et~al.}(2023){Cullen}, {McLure}, {McLeod}, {Dunlop},
  {Donnan}, {Carnall}, {Bowler}, {Begley}, {Hamadouche}, \&
  {Stanton}}]{Cullen2023}
{Cullen}, F., {McLure}, R.~J., {McLeod}, D.~J., {et~al.} 2023, \mnras, 520, 14,
  \dodoi{10.1093/mnras/stad073}

\bibitem[{{Diego} {et~al.}(2023){Diego}, {Adams}, {Willner}, {Harvey},
  {Broadhurst}, {Cohen}, {Jansen}, {Summers}, {Windhorst}, {D'Silva},
  {Koekemoer}, {Coe}, {Conselice}, {Driver}, {Frye}, {Grogin}, {Marshall},
  {Nonino}, {Ortiz}, {Pirzkal}, {Robotham}, {Ryan}, {Willmer}, {Yan}, {Sun},
  {Hainline}, {Berkheimer}, {Polletta}, \& {Zitrin}}]{Diego2023}
{Diego}, J.~M., {Adams}, N.~J., {Willner}, S., {et~al.} 2023, arXiv e-prints,
  arXiv:2312.11603, \dodoi{10.48550/arXiv.2312.11603}

\bibitem[{{Dunlop} {et~al.}(2012){Dunlop}, {McLure}, {Robertson}, {Ellis},
  {Stark}, {Cirasuolo}, \& {de Ravel}}]{Dunlop2012}
{Dunlop}, J.~S., {McLure}, R.~J., {Robertson}, B.~E., {et~al.} 2012, \mnras,
  420, 901, \dodoi{10.1111/j.1365-2966.2011.20102.x}

\bibitem[{{Dunlop} {et~al.}(2013){Dunlop}, {Rogers}, {McLure}, {Ellis},
  {Robertson}, {Koekemoer}, {Dayal}, {Curtis-Lake}, {Wild}, {Charlot},
  {Bowler}, {Schenker}, {Ouchi}, {Ono}, {Cirasuolo}, {Furlanetto}, {Stark},
  {Targett}, \& {Schneider}}]{Dunlop2013}
{Dunlop}, J.~S., {Rogers}, A.~B., {McLure}, R.~J., {et~al.} 2013, \mnras, 432,
  3520, \dodoi{10.1093/mnras/stt702}

\bibitem[{{Finkelstein} {et~al.}(2012){Finkelstein}, {Papovich}, {Salmon},
  {Finlator}, {Dickinson}, {Ferguson}, {Giavalisco}, {Koekemoer}, {Reddy},
  {Bassett}, {Conselice}, {Dunlop}, {Faber}, {Grogin}, {Hathi}, {Kocevski},
  {Lai}, {Lee}, {McLure}, {Mobasher}, \& {Newman}}]{Finkelstein2012}
{Finkelstein}, S.~L., {Papovich}, C., {Salmon}, B., {et~al.} 2012, \apj, 756,
  164, \dodoi{10.1088/0004-637X/756/2/164}

\bibitem[{{Flury} {et~al.}(2022){Flury}, {Jaskot}, {Ferguson}, {Worseck},
  {Makan}, {Chisholm}, {Saldana-Lopez}, {Schaerer}, {McCandliss}, {Wang},
  {Ford}, {Heckman}, {Ji}, {Giavalisco}, {Amorin}, {Atek}, {Blaizot},
  {Borthakur}, {Carr}, {Castellano}, {Cristiani}, {De Barros}, {Dickinson},
  {Finkelstein}, {Fleming}, {Fontanot}, {Garel}, {Grazian}, {Hayes}, {Henry},
  {Mauerhofer}, {Micheva}, {Oey}, {Ostlin}, {Papovich}, {Pentericci},
  {Ravindranath}, {Rosdahl}, {Rutkowski}, {Santini}, {Scarlata}, {Teplitz},
  {Thuan}, {Trebitsch}, {Vanzella}, {Verhamme}, \& {Xu}}]{Flury2022}
{Flury}, S.~R., {Jaskot}, A.~E., {Ferguson}, H.~C., {et~al.} 2022, \apjs, 260,
  1, \dodoi{10.3847/1538-4365/ac5331}

\bibitem[{{Fujimoto} {et~al.}(2023){Fujimoto}, {Arrabal Haro}, {Dickinson},
  {Finkelstein}, {Kartaltepe}, {Larson}, {Burgarella}, {Bagley}, {Behroozi},
  {Chworowsky}, {Hirschmann}, {Trump}, {Wilkins}, {Yung}, {Koekemoer},
  {Papovich}, {Pirzkal}, {Ferguson}, {Fontana}, {Grogin}, {Grazian}, {Kewley},
  {Kocevski}, {Lotz}, {Pentericci}, {Ravindranath}, {Somerville}, {Wilkins},
  {Amor{\'\i}n}, {Backhaus}, {Calabr{\`o}}, {Casey}, {Cooper}, {Fern{\'a}ndez},
  {Franco}, {Giavalisco}, {Hathi}, {Harish}, {Hutchison}, {Iyer}, {Jung},
  {Lucas}, \& {Zavala}}]{Fujimoto2023a}
{Fujimoto}, S., {Arrabal Haro}, P., {Dickinson}, M., {et~al.} 2023, \apjl, 949,
  L25, \dodoi{10.3847/2041-8213/acd2d9}

\bibitem[{{Grazian} {et~al.}(2016){Grazian}, {Giallongo}, {Gerbasi}, {Fiore},
  {Fontana}, {Le F{\`e}vre}, {Pentericci}, {Vanzella}, {Zamorani}, {Cassata},
  {Garilli}, {Le Brun}, {Maccagni}, {Tasca}, {Thomas}, {Zucca}, {Amor{\'\i}n},
  {Bardelli}, {Cassar{\`a}}, {Castellano}, {Cimatti}, {Cucciati}, {Durkalec},
  {Giavalisco}, {Hathi}, {Ilbert}, {Lemaux}, {Paltani}, {Ribeiro}, {Schaerer},
  {Scodeggio}, {Sommariva}, {Talia}, {Tresse}, {Vergani}, {Bonchi}, {Boutsia},
  {Capak}, {Charlot}, {Contini}, {de la Torre}, {Dunlop}, {Fotopoulou},
  {Guaita}, {Koekemoer}, {L{\'o}pez-Sanjuan}, {Mellier}, {Merlin}, {Paris},
  {Pforr}, {Pilo}, {Santini}, {Scoville}, {Taniguchi}, \& {Wang}}]{Grazian2016}
{Grazian}, A., {Giallongo}, E., {Gerbasi}, R., {et~al.} 2016, \aap, 585, A48,
  \dodoi{10.1051/0004-6361/201526396}

\bibitem[{{Griffiths} {et~al.}(2022){Griffiths}, {Conselice}, {Ferreira},
  {Ceverino}, {P{\'e}rez-Gonz{\'a}lez}, {Vega}, {Rosa-Gonz{\'a}lez},
  {Koekemoer}, {Marchesini}, {Rodr{\'\i}guez Espinosa},
  {Rodr{\'\i}guez-Mu{\~n}oz}, {Pampliega}, \& {Terlevich}}]{griffiths2022}
{Griffiths}, A., {Conselice}, C.~J., {Ferreira}, L., {et~al.} 2022, \apj, 941,
  181, \dodoi{10.3847/1538-4357/aca296}

\bibitem[{{Harikane} {et~al.}(2024){Harikane}, {Nakajima}, {Ouchi}, {Umeda},
  {Isobe}, {Ono}, {Xu}, \& {Zhang}}]{Harikane2024}
{Harikane}, Y., {Nakajima}, K., {Ouchi}, M., {et~al.} 2024, \apj, 960, 56,
  \dodoi{10.3847/1538-4357/ad0b7e}

\bibitem[{{Hathi} {et~al.}(2008){Hathi}, {Malhotra}, \& {Rhoads}}]{Hathi2008}
{Hathi}, N.~P., {Malhotra}, S., \& {Rhoads}, J.~E. 2008, \apj, 673, 686,
  \dodoi{10.1086/524836}

\bibitem[{{Helton} {et~al.}(2023){Helton}, {Sun}, {Woodrum}, {Hainline},
  {Willmer}, {Rieke}, {Rieke}, {Alberts}, {Eisenstein}, {Tacchella},
  {Robertson}, {Johnson}, {Baker}, {Bhatawdekar}, {Bunker}, {Chen}, {Egami},
  {Ji}, {Maiolino}, {Willott}, \& {Witstok}}]{Helton2023}
{Helton}, J.~M., {Sun}, F., {Woodrum}, C., {et~al.} 2023, arXiv e-prints,
  arXiv:2311.04270, \dodoi{10.48550/arXiv.2311.04270}

\bibitem[{{Hsiao} {et~al.}(2023){Hsiao}, {Abdurro'uf}, {Coe}, {Larson}, {Jung},
  {Mingozzi}, {Dayal}, {Kumari}, {Kokorev}, {Vikaeus}, {Brammer}, {Furtak},
  {Adamo}, {Andrade-Santos}, {Antwi-Danso}, {Bradac}, {Bradley}, {Broadhurst},
  {Carnall}, {Conselice}, {Diego}, {Donahue}, {Eldridge}, {Fujimoto}, {Henry},
  {Hernandez}, {Hutchison}, {James}, {Norman}, {Park}, {Pirzkal}, {Postman},
  {Ricotti}, {Rigby}, {Vanzella}, {Welch}, {Wilkins}, {Windhorst}, {Xu},
  {Zackrisson}, \& {Zitrin}}]{Hsiao2023}
{Hsiao}, T. Y.-Y., {Abdurro'uf}, {Coe}, D., {et~al.} 2023, arXiv e-prints,
  arXiv:2305.03042, \dodoi{10.48550/arXiv.2305.03042}

\bibitem[{{Infante} {et~al.}(2015){Infante}, {Zheng}, {Laporte}, {Troncoso
  Iribarren}, {Molino}, {Diego}, {Bauer}, {Zitrin}, {Moustakas}, {Huang},
  {Shu}, {Bina}, {Brammer}, {Broadhurst}, {Ford}, {Garc{\'\i}a}, \&
  {Kim}}]{Infante2015}
{Infante}, L., {Zheng}, W., {Laporte}, N., {et~al.} 2015, \apj, 815, 18,
  \dodoi{10.1088/0004-637X/815/1/18}

\bibitem[{{Izotov} {et~al.}(2016){Izotov}, {Schaerer}, {Thuan}, {Worseck},
  {Guseva}, {Orlitov{\'a}}, \& {Verhamme}}]{Izotov2016}
{Izotov}, Y.~I., {Schaerer}, D., {Thuan}, T.~X., {et~al.} 2016, \mnras, 461,
  3683, \dodoi{10.1093/mnras/stw1205}

\bibitem[{{Izotov} {et~al.}(2021){Izotov}, {Worseck}, {Schaerer}, {Guseva},
  {Chisholm}, {Thuan}, {Fricke}, \& {Verhamme}}]{Izotov2021}
{Izotov}, Y.~I., {Worseck}, G., {Schaerer}, D., {et~al.} 2021, \mnras, 503,
  1734, \dodoi{10.1093/mnras/stab612}

\bibitem[{{Izotov} {et~al.}(2018){Izotov}, {Worseck}, {Schaerer}, {Guseva},
  {Thuan}, {Fricke}, \& {Orlitov{\'a}}}]{Izotov2018}
---. 2018, \mnras, 478, 4851, \dodoi{10.1093/mnras/sty1378}

\bibitem[{{Kashino} {et~al.}(2023){Kashino}, {Lilly}, {Matthee}, {Eilers},
  {Mackenzie}, {Bordoloi}, \& {Simcoe}}]{Kashino2023}
{Kashino}, D., {Lilly}, S.~J., {Matthee}, J., {et~al.} 2023, \apj, 950, 66,
  \dodoi{10.3847/1538-4357/acc588}

\bibitem[{{Kokorev} {et~al.}(2022){Kokorev}, {Brammer}, {Fujimoto}, {Kohno},
  {Magdis}, {Valentino}, {Toft}, {Oesch}, {Davidzon}, {Bauer}, {Coe}, {Egami},
  {Oguri}, {Ouchi}, {Postman}, {Richard}, {Jolly}, {Knudsen}, {Sun}, {Weaver},
  {Ao}, {Baker}, {Bradley}, {Caputi}, {Dessauges-Zavadsky}, {Espada},
  {Hatsukade}, {Koekemoer}, {Mu{\~n}oz Arancibia}, {Shimasaku}, {Umehata},
  {Wang}, \& {Wang}}]{Kokorev2022}
{Kokorev}, V., {Brammer}, G., {Fujimoto}, S., {et~al.} 2022, \apjs, 263, 38,
  \dodoi{10.3847/1538-4365/ac9909}

\bibitem[{{Kroupa}(2001)}]{Kroupa2001}
{Kroupa}, P. 2001, \mnras, 322, 231, \dodoi{10.1046/j.1365-8711.2001.04022.x}

\bibitem[{{Laporte} {et~al.}(2021){Laporte}, {Meyer}, {Ellis}, {Robertson},
  {Chisholm}, \& {Roberts-Borsani}}]{Laporte2021MNRAS}
{Laporte}, N., {Meyer}, R.~A., {Ellis}, R.~S., {et~al.} 2021, \mnras, 505,
  3336, \dodoi{10.1093/mnras/stab1239}

\bibitem[{{Laporte} {et~al.}(2015){Laporte}, {Streblyanska}, {Kim},
  {Pell{\'o}}, {Bauer}, {Bina}, {Brammer}, {De Leo}, {Infante}, \&
  {P{\'e}rez-Fournon}}]{Laporte2015}
{Laporte}, N., {Streblyanska}, A., {Kim}, S., {et~al.} 2015, \aap, 575, A92,
  \dodoi{10.1051/0004-6361/201425040}

\bibitem[{{Laporte} {et~al.}(2016){Laporte}, {Infante}, {Troncoso Iribarren},
  {Zheng}, {Molino}, {Bauer}, {Bina}, {Broadhurst}, {Chilingarian}, {Huang},
  {Garcia}, {Kim}, {Marques-Chaves}, {Moustakas}, {Pell{\'o}},
  {P{\'e}rez-Fournon}, {Shu}, {Streblyanska}, \& {Zitrin}}]{Laporte2016}
{Laporte}, N., {Infante}, L., {Troncoso Iribarren}, P., {et~al.} 2016, \apj,
  820, 98, \dodoi{10.3847/0004-637X/820/2/98}

\bibitem[{{Lotz} {et~al.}(2017){Lotz}, {Koekemoer}, {Coe}, {Grogin}, {Capak},
  {Mack}, {Anderson}, {Avila}, {Barker}, {Borncamp}, {Brammer}, {Durbin},
  {Gunning}, {Hilbert}, {Jenkner}, {Khandrika}, {Levay}, {Lucas}, {MacKenty},
  {Ogaz}, {Porterfield}, {Reid}, {Robberto}, {Royle}, {Smith},
  {Storrie-Lombardi}, {Sunnquist}, {Surace}, {Taylor}, {Williams}, {Bullock},
  {Dickinson}, {Finkelstein}, {Natarajan}, {Richard}, {Robertson}, {Tumlinson},
  {Zitrin}, {Flanagan}, {Sembach}, {Soifer}, \& {Mountain}}]{Lotz2017}
{Lotz}, J.~M., {Koekemoer}, A., {Coe}, D., {et~al.} 2017, \apj, 837, 97,
  \dodoi{10.3847/1538-4357/837/1/97}

\bibitem[{{Morales} {et~al.}(2024){Morales}, {Finkelstein}, {Leung}, {Bagley},
  {Cleri}, {Dave}, {Dickinson}, {Ferguson}, {Hathi}, {Jones}, {Koekemoer},
  {Papovich}, {P{\'e}rez-Gonz{\'a}lez}, {Pirzkal}, {Smith}, {Wilkins}, \&
  {Yung}}]{Morales2024}
{Morales}, A.~M., {Finkelstein}, S.~L., {Leung}, G. C.~K., {et~al.} 2024,
  \apjl, 964, L24, \dodoi{10.3847/2041-8213/ad2de4}

\bibitem[{{Nanayakkara} {et~al.}(2023){Nanayakkara}, {Glazebrook}, {Jacobs},
  {Bonchi}, {Castellano}, {Fontana}, {Mason}, {Merlin}, {Morishita}, {Paris},
  {Trenti}, {Treu}, {Calabr{\`o}}, {Boyett}, {Bradac}, {Leethochawalit},
  {Marchesini}, {Santini}, {Strait}, {Vanzella}, {Vulcani}, {Wang}, \&
  {Yang}}]{Nanayakkara2023}
{Nanayakkara}, T., {Glazebrook}, K., {Jacobs}, C., {et~al.} 2023, \apjl, 947,
  L26, \dodoi{10.3847/2041-8213/acbfb9}

\bibitem[{{Peng} {et~al.}(2002){Peng}, {Ho}, {Impey}, \& {Rix}}]{Peng2002}
{Peng}, C.~Y., {Ho}, L.~C., {Impey}, C.~D., \& {Rix}, H.-W. 2002, \aj, 124,
  266, \dodoi{10.1086/340952}

\bibitem[{{Peng} {et~al.}(2010){Peng}, {Ho}, {Impey}, \& {Rix}}]{Peng2010}
---. 2010, \aj, 139, 2097, \dodoi{10.1088/0004-6256/139/6/2097}

\bibitem[{{Perrin} {et~al.}(2015){Perrin}, {Long}, {Sivaramakrishnan},
  {Lajoie}, {Elliot}, {Pueyo}, \& {Albert}}]{Perrin2015}
{Perrin}, M.~D., {Long}, J., {Sivaramakrishnan}, A., {et~al.} 2015, {WebbPSF:
  James Webb Space Telescope PSF Simulation Tool}, Astrophysics Source Code
  Library, record ascl:1504.007.
\newblock \doeprint{1504.007}

\bibitem[{{Perrin} {et~al.}(2014){Perrin}, {Sivaramakrishnan}, {Lajoie},
  {Elliott}, {Pueyo}, {Ravindranath}, \& {Albert}}]{Perrin2014}
{Perrin}, M.~D., {Sivaramakrishnan}, A., {Lajoie}, C.-P., {et~al.} 2014, in
  Society of Photo-Optical Instrumentation Engineers (SPIE) Conference Series,
  Vol. 9143, Space Telescopes and Instrumentation 2014: Optical, Infrared, and
  Millimeter Wave, ed. J.~{Oschmann}, Jacobus~M., M.~{Clampin}, G.~G. {Fazio},
  \& H.~A. {MacEwen}, 91433X, \dodoi{10.1117/12.2056689}

\bibitem[{{Popping}(2023)}]{Popping2023}
{Popping}, G. 2023, \aap, 669, L8, \dodoi{10.1051/0004-6361/202244831}

\bibitem[{{Roberts-Borsani} {et~al.}(2024){Roberts-Borsani}, {Treu}, {Shapley},
  {Fontana}, {Pentericci}, {Castellano}, {Morishita}, {Bergamini}, \&
  {Rosati}}]{RB2024}
{Roberts-Borsani}, G., {Treu}, T., {Shapley}, A., {et~al.} 2024, arXiv
  e-prints, arXiv:2403.07103, \dodoi{10.48550/arXiv.2403.07103}

\bibitem[{{Robertson}(2022)}]{Robertson2022}
{Robertson}, B.~E. 2022, \araa, 60, 121,
  \dodoi{10.1146/annurev-astro-120221-044656}

\bibitem[{{Saldana-Lopez} {et~al.}(2022){Saldana-Lopez}, {Schaerer},
  {Chisholm}, {Flury}, {Jaskot}, {Worseck}, {Makan}, {Gazagnes}, {Mauerhofer},
  {Verhamme}, {Amor{\'\i}n}, {Ferguson}, {Giavalisco}, {Grazian}, {Hayes},
  {Heckman}, {Henry}, {Ji}, {Marques-Chaves}, {McCandliss}, {Oey},
  {{\"O}stlin}, {Pentericci}, {Thuan}, {Trebitsch}, {Vanzella}, \&
  {Xu}}]{SL2022}
{Saldana-Lopez}, A., {Schaerer}, D., {Chisholm}, J., {et~al.} 2022, \aap, 663,
  A59, \dodoi{10.1051/0004-6361/202141864}

\bibitem[{{Saxena} {et~al.}(2024){Saxena}, {Bunker}, {Jones}, {Stark},
  {Cameron}, {Witstok}, {Arribas}, {Baker}, {Baum}, {Bhatawdekar}, {Bowler},
  {Boyett}, {Carniani}, {Charlot}, {Chevallard}, {Curti}, {Curtis-Lake},
  {Eisenstein}, {Endsley}, {Hainline}, {Helton}, {Johnson}, {Kumari}, {Looser},
  {Maiolino}, {Rieke}, {Rix}, {Robertson}, {Sandles}, {Simmonds}, {Smit},
  {Tacchella}, {Williams}, {Willmer}, \& {Willott}}]{Saxena2024}
{Saxena}, A., {Bunker}, A.~J., {Jones}, G.~C., {et~al.} 2024, \aap, 684, A84,
  \dodoi{10.1051/0004-6361/202347132}

\bibitem[{{S{\'e}rsic}(1963)}]{Sersic1963}
{S{\'e}rsic}, J.~L. 1963, Boletin de la Asociacion Argentina de Astronomia La
  Plata Argentina, 6, 41

\bibitem[{{Siana} {et~al.}(2015){Siana}, {Shapley}, {Kulas}, {Nestor},
  {Steidel}, {Teplitz}, {Alavi}, {Brown}, {Conselice}, {Ferguson}, {Dickinson},
  {Giavalisco}, {Colbert}, {Bridge}, {Gardner}, \& {de Mello}}]{Siana2015}
{Siana}, B., {Shapley}, A.~E., {Kulas}, K.~R., {et~al.} 2015, \apj, 804, 17,
  \dodoi{10.1088/0004-637X/804/1/17}

\bibitem[{{Steidel} {et~al.}(2018){Steidel}, {Bogosavljevi{\'c}}, {Shapley},
  {Reddy}, {Rudie}, {Pettini}, {Trainor}, \& {Strom}}]{Steidel2018}
{Steidel}, C.~C., {Bogosavljevi{\'c}}, M., {Shapley}, A.~E., {et~al.} 2018,
  \apj, 869, 123, \dodoi{10.3847/1538-4357/aaed28}

\bibitem[{{Steidel} {et~al.}(2001){Steidel}, {Pettini}, \&
  {Adelberger}}]{Steidel2001}
{Steidel}, C.~C., {Pettini}, M., \& {Adelberger}, K.~L. 2001, \apj, 546, 665,
  \dodoi{10.1086/318323}

\bibitem[{{Sun} {et~al.}(2023){Sun}, {Egami}, {Pirzkal}, {Rieke}, {Baum},
  {Boyer}, {Boyett}, {Bunker}, {Cameron}, {Curti}, {Eisenstein}, {Gennaro},
  {Greene}, {Jaffe}, {Kelly}, {Koekemoer}, {Kumari}, {Maiolino}, {Maseda},
  {Perna}, {Rest}, {Robertson}, {Schlawin}, {Smit}, {Stansberry}, {Sunnquist},
  {Tacchella}, {Williams}, \& {Willmer}}]{Sun2023_wfss}
{Sun}, F., {Egami}, E., {Pirzkal}, N., {et~al.} 2023, \apj, 953, 53,
  \dodoi{10.3847/1538-4357/acd53c}

\bibitem[{{Tamura} {et~al.}(2019){Tamura}, {Mawatari}, {Hashimoto}, {Inoue},
  {Zackrisson}, {Christensen}, {Binggeli}, {Matsuda}, {Matsuo}, {Takeuchi},
  {Asano}, {Sunaga}, {Shimizu}, {Okamoto}, {Yoshida}, {Lee}, {Shibuya},
  {Taniguchi}, {Umehata}, {Hatsukade}, {Kohno}, \& {Ota}}]{Tamura2019}
{Tamura}, Y., {Mawatari}, K., {Hashimoto}, T., {et~al.} 2019, \apj, 874, 27,
  \dodoi{10.3847/1538-4357/ab0374}

\bibitem[{{Tamura} {et~al.}(2023){Tamura}, {C. Bakx}, {Inoue}, {Hashimoto},
  {Tokuoka}, {Imamura}, {Hatsukade}, {Lee}, {Moriwaki}, {Okamoto}, {Ota},
  {Umehata}, {Yoshida}, {Zackrisson}, {Hagimoto}, {Matsuo}, {Shimizu},
  {Sugahara}, \& {Takeuchi}}]{Tamura2023}
{Tamura}, Y., {C. Bakx}, T. J.~L., {Inoue}, A.~K., {et~al.} 2023, \apj, 952, 9,
  \dodoi{10.3847/1538-4357/acd637}

\bibitem[{{Tang} {et~al.}(2023){Tang}, {Stark}, {Chen}, {Mason}, {Topping},
  {Endsley}, {Senchyna}, {Plat}, {Lu}, {Whitler}, {Robertson}, \&
  {Charlot}}]{Tang2023}
{Tang}, M., {Stark}, D.~P., {Chen}, Z., {et~al.} 2023, \mnras, 526, 1657,
  \dodoi{10.1093/mnras/stad2763}

\bibitem[{{Topping} {et~al.}(2022){Topping}, {Stark}, {Endsley}, {Plat},
  {Whitler}, {Chen}, \& {Charlot}}]{Topping2022}
{Topping}, M.~W., {Stark}, D.~P., {Endsley}, R., {et~al.} 2022, \apj, 941, 153,
  \dodoi{10.3847/1538-4357/aca522}

\bibitem[{{Vanzella} {et~al.}(2012){Vanzella}, {Guo}, {Giavalisco}, {Grazian},
  {Castellano}, {Cristiani}, {Dickinson}, {Fontana}, {Nonino}, {Giallongo},
  {Pentericci}, {Galametz}, {Faber}, {Ferguson}, {Grogin}, {Koekemoer},
  {Newman}, \& {Siana}}]{Vanzella2012}
{Vanzella}, E., {Guo}, Y., {Giavalisco}, M., {et~al.} 2012, \apj, 751, 70,
  \dodoi{10.1088/0004-637X/751/1/70}

\bibitem[{{Windhorst} {et~al.}(2023){Windhorst}, {Cohen}, {Jansen}, {Summers},
  {Tompkins}, {Conselice}, {Driver}, {Yan}, {Coe}, {Frye}, {Grogin},
  {Koekemoer}, {Marshall}, {O'Brien}, {Pirzkal}, {Robotham}, {Ryan}, {Willmer},
  {Carleton}, {Diego}, {Keel}, {Porto}, {Redshaw}, {Scheller}, {Wilkins},
  {Willner}, {Zitrin}, {Adams}, {Austin}, {Arendt}, {Beacom}, {Bhatawdekar},
  {Bradley}, {Broadhurst}, {Cheng}, {Civano}, {Dai}, {Dole}, {D'Silva},
  {Duncan}, {Fazio}, {Ferrami}, {Ferreira}, {Finkelstein}, {Furtak}, {Gim},
  {Griffiths}, {Hammel}, {Harrington}, {Hathi}, {Holwerda}, {Honor}, {Huang},
  {Hyun}, {Im}, {Joshi}, {Kamieneski}, {Kelly}, {Larson}, {Li}, {Lim}, {Ma},
  {Maksym}, {Manzoni}, {Meena}, {Milam}, {Nonino}, {Pascale}, {Petric},
  {Pierel}, {del Carmen Polletta}, {R{\"o}ttgering}, {Rutkowski}, {Smail},
  {Straughn}, {Strolger}, {Swirbul}, {Trussler}, {Wang}, {Welch}, {B. Wyithe},
  {Yun}, {Zackrisson}, {Zhang}, \& {Zhao}}]{Windhorst2023}
{Windhorst}, R.~A., {Cohen}, S.~H., {Jansen}, R.~A., {et~al.} 2023, \aj, 165,
  13, \dodoi{10.3847/1538-3881/aca163}

\bibitem[{{Yan} \& {Windhorst}(2004)}]{Yan2004a}
{Yan}, H., \& {Windhorst}, R.~A. 2004, \apjl, 600, L1, \dodoi{10.1086/381573}

\bibitem[{{Yan} {et~al.}(2023){Yan}, {Ma}, {Sun}, {Wang}, {Kelly}, {Diego},
  {Cohen}, {Windhorst}, {Jansen}, {Grogin}, {Beacom}, {Conselice}, {Driver},
  {Frye}, {Coe}, {Marshall}, {Koekemoer}, {Willmer}, {Robotham}, {D'Silva},
  {Summers}, {Nonino}, {Pirzkal}, {Ryan}, {Ortiz}, {Tompkins}, {Bhatawdekar},
  {Cheng}, {Zitrin}, \& {Willner}}]{Yan2023c}
{Yan}, H., {Ma}, Z., {Sun}, B., {et~al.} 2023, \apjs, 269, 43,
  \dodoi{10.3847/1538-4365/ad0298}

\bibitem[{{Zackrisson} {et~al.}(2013){Zackrisson}, {Inoue}, \&
  {Jensen}}]{Zackrisson2013}
{Zackrisson}, E., {Inoue}, A.~K., \& {Jensen}, H. 2013, \apj, 777, 39,
  \dodoi{10.1088/0004-637X/777/1/39}

\end{thebibliography}
\bibliographystyle{aasjournal}

\appendix
\counterwithin{figure}{section}
\counterwithin{table}{section}

{\section{$z_{\rm ph}$ of f090d\_018 and identification of the [\ion{O}{3}] line}

    The new object f090d\_018 is among the F090W dropouts that we selected in this 
field and passes our photometric-redshift test as a legitimate $z>8$ candidate. 
Running \textsc{Bagpipes} with redshift as a free parameter gives 
$z_{\rm ph}=8.76^{+0.04}_{-0.04}$, and Figure~\ref{fig:f090d_zph} shows the results.

    As mentioned in Section 3.3 and shown in Figure~\ref{fig:spec}, only one emission
line, at 4.752~$\mu$m,  was detected for f090d\_018.  Within the wide range of 
$7.5<z<9.5$, the line could be either H$\beta$~$\lambda$4861 at $z=8.775$ or 
[\ion{O}{3}]~$\lambda$5007 at $z=8.490$. If it is the former, 
the [\ion{O}{3}]~$\lambda$5007 line, which is usually stronger
than H$\beta$~$\lambda$4861, would be observed at 4.894~$\mu$m,
well within the high-sensitivity range of the same spectrum. However, no
line is detected at this wavelength. Unless this galaxy has a highly abnormal 
abundance (e.g., close to being made of only Pop~\Romannum{3} stars), this null
detection would be difficult to explain.
Therefore, we are left with the line being 
[\ion{O}{3}]~$\lambda$5007. The difference between $z_{\rm ph}$ and the spectroscopic
redshift thus determined is $\Delta z=z_{\rm ph}-z =0.27$, and 
$\Delta z/(1+z)=0.028$, which is reasonable.

\begin{figure}
    \centering
    \includegraphics[width=0.55\linewidth]{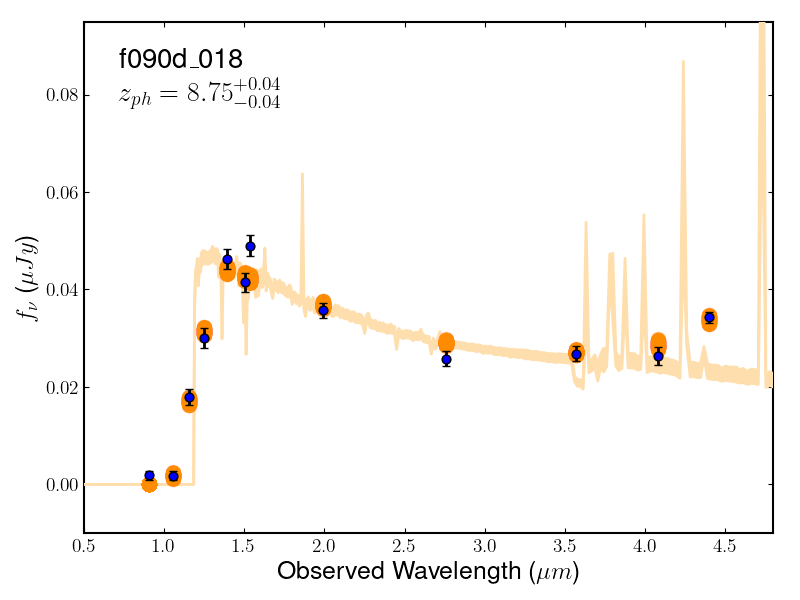}
    \includegraphics[width=0.8\linewidth]{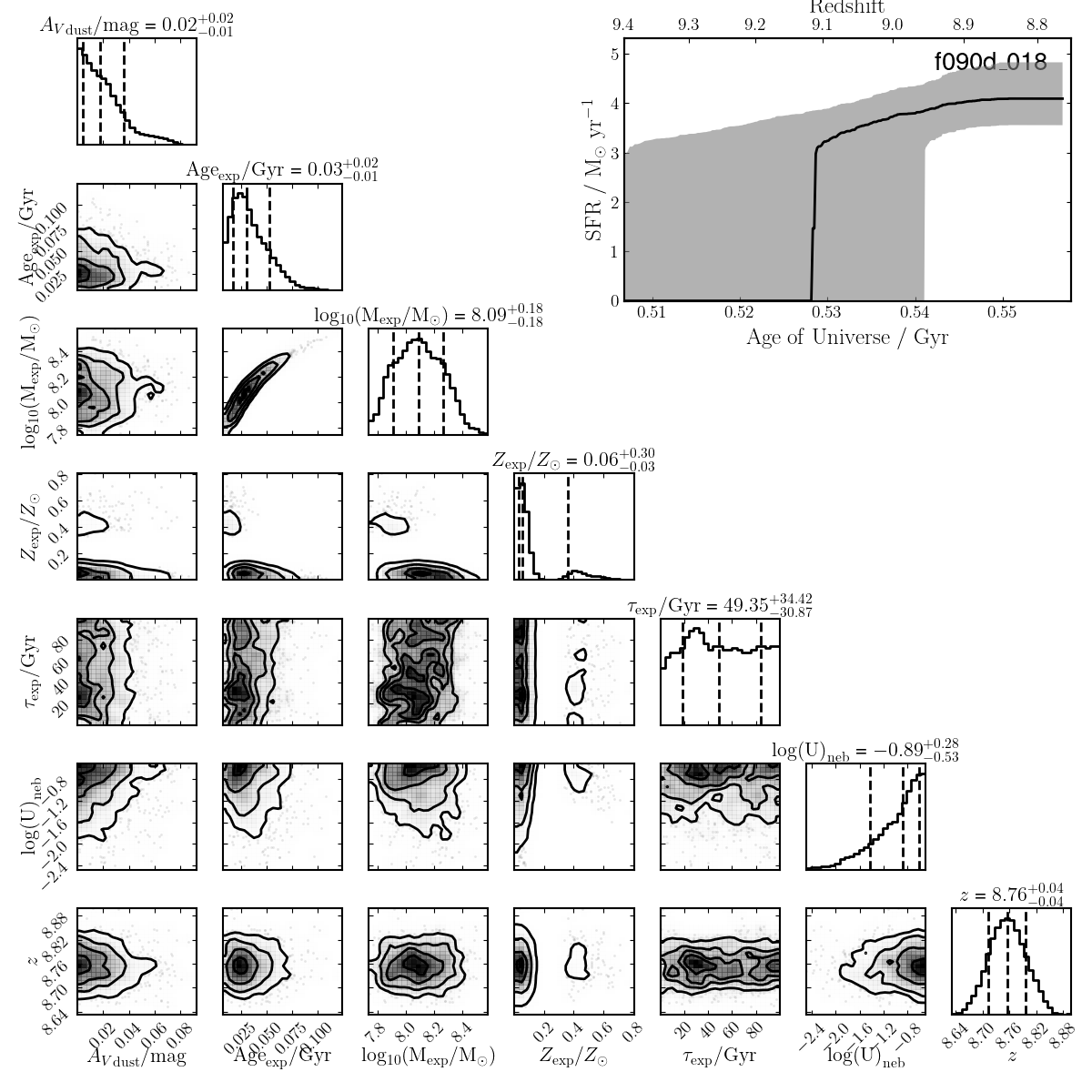}
    \caption{SED fitting results for f090d\_018 with redshift as a free parameter to derive $z_{\rm ph}$. The plots are similar to those in Figures~\ref{fig:sed_fitting} and \ref{fig:sed_fitting_corner}.}
    \label{fig:f090d_zph}
\end{figure}

}

\section{Age estimates using different SFHs}

    For a given stellar population synthesis model, the age estimate could be
affected by the adopted SFH\null. The SED analysis presented in Section 
\ref{sec:bagpipes} uses an exponentially declining SFH, also known as the 
``$\tau$ model''. To test the robustness of the very young ages thus obtained, 
we also ran \textsc{Bagpipes} using three other SFHs offered in the package:
\begin{itemize}
    %\item exponential decay. See \S~{\ref{sec:bagpipes}}.
    \item Delayed $\tau$:
\begin{equation}\label{eqn:del}
\quad \mathrm{SFR}(t)\ \propto\  te^{-t/\tau}
\end{equation}
    \item Log-normal: 
\begin{equation}\label{eqn:lognorm}
\mathrm{SFR}(t)\ \propto\ \frac{1}{t}\ \mathrm{exp} \bigg( -\frac{\big(\mathrm{ln}(t)\big)^2}{2\tau^2} \bigg)
\end{equation}
    \item Double power law:
\begin{equation}\label{eqn:DPL}
\mathrm{SFR}(t)\ \propto\ \Bigg[\bigg(\frac{t}{\tau}\bigg)^{\alpha} + \bigg(\frac{t}{\tau}\bigg)^{-\beta}\Bigg]^{-1}
\end{equation}
\end{itemize}

    Table~\ref{tbl:sfh_check} compares the best-fit stellar mass and age 
parameters obtained using the $\tau$ model with those obtained using these 
alternative SFHs. For both the $\tau$ and the delayed $\tau$ models, the age
parameter is among the direct outputs. This is not the case when using the 
log-normal or the double power law models, however, because there is no clear
definition of age in either SFH. To obtain an estimate in these two cases, we 
derived an age proxy denoted as ${\rm Age_{N}}$, which is the time for
the galaxy to gain 100\% of its stellar mass starting from the time when it had 
N\% of its total stellar mass. For the demonstration purpose here, $N$ was set to 10, 50, and 90. As shown by this comparison, these alternative SFHs 
resulted in comparable, young ages for all three objects. 

%The results are summarized in Table~\ref{tbl:sfh_check}

\begin{table*}
\centering
\caption{Comparison of best-fit galaxy properties with different SFH}
\begin{tabular}{lcccccc}
\hline\hline
   & & Exponential & Delayed $\tau$ & Log-norm & Double PL \\
\hline
Y1 & ${\rm log}(\mu M/\mathrm{M_{\sun}})$ & $8.97^{+0.03}_{-0.01}$ & $8.90^{+0.07}_{-0.10}$ & $8.84^{+0.05}_{-0.07}$ & $8.85^{+0.05}_{-0.07}$ \\
       & Age/Myr & $4.76^{+0.28}_{-0.35}$ & $3.51^{+1.68}_{-1.82}$ & ... & ... \\
       & Age$_{10}$/Myr & ... & ... & 4.10 & 3.33 \\
       & Age$_{50}$/Myr & ... & ... & 1.39 & 1.29 \\
       & Age$_{90}$/Myr & ... & ... & 1.00 & 1.00 \\
\hline
JD & ${\rm log}(\mu M/\mathrm{M_{\sun}})$ &  $8.12^{+0.12}_{-0.11}$ & $8.13^{+0.18}_{-0.11}$ & $8.13^{+0.09}_{-0.10}$ & $8.15^{+0.08}_{-0.08}$ \\
       & Age/Myr & $4.28^{+1.06}_{-1.17}$ & $5.89^{+2.36}_{-2.60}$ & ... & ... \\   
       & Age$_{10}$/Myr & ... & ... & 6.42 & 5.13 \\
       & Age$_{50}$/Myr & ... & ... & 2.05 & 2.01 \\
       & Age$_{90}$/Myr & ... & ... & 1.00 & 1.00 \\      
\hline
JD-N & ${\rm log}(\mu M/\mathrm{M_{\sun}})$ & $8.95^{+0.05}_{-0.04}$ & $8.99^{+0.05}_{-0.06}$ & $9.02^{+0.06}_{-0.06}$ & $8.99^{+0.05}_{-0.05}$ \\
       & Age/Myr & $8.69^{+2.46}_{-2.19}$ & $12.77^{+7.24}_{-4.42}$ & ... & ... \\
       & Age$_{10}$/Myr & ... & ... & 10.86 & 7.85 \\
       & Age$_{50}$/Myr & ... & ... & 3.36 & 2.82 \\
       & Age$_{90}$/Myr & ... & ... & 1.00 & 1.00 \\      
\hline
f090d\_018 & ${\rm log}(\mu M/\mathrm{M_{\sun}})$ & $7.88^{+0.19}_{-0.17}$ & $7.89^{+0.21}_{-0.17}$ & $7.84^{+0.18}_{-0.13}$ & $7.91^{+0.21}_{-0.17}$ \\
       & Age/Myr & $17.79^{+15.50}_{-7.56}$ & $33.11^{+37.20}_{-15.59}$ & ... & ... \\
       & Age$_{10}$/Myr & ... & ... & 21.50 & 15.84 \\
       & Age$_{50}$/Myr & ... & ... & 6.62 & 8.81 \\
       & Age$_{90}$/Myr & ... & ... & 1.05 & 1.81 \\      
\hline
\end{tabular}
\raggedright
\tablecomments{The ages for the log-normal and double power law SFHs are not
available from \textsc{Bagpipes} direct outputs because the parameter has no
clear meaning in either SFH. We define ${\rm Age_{N}}$ for these two cases, 
where ${\rm N=10}$, 50, and 90. See text for details.
}\label{tbl:sfh_check}
\end{table*}

\end{document}